%% file: V19_Else_V03.tex
\documentclass[5p]{elsarticle}

\usepackage{amsmath,amsfonts,amssymb}
\usepackage{algorithmic}
\usepackage{algorithm}
\usepackage{array}
\usepackage{textcomp}
\usepackage{url}
\usepackage{verbatim}
\usepackage{graphicx}
\usepackage{booktabs}
\usepackage{xcolor}
\usepackage{colortbl}
\usepackage{tabularx}
\usepackage{makecell}
\usepackage{multirow}
\usepackage{lineno}
\usepackage{subfig}

\usepackage{tikz}
\usetikzlibrary{positioning, shapes.geometric, arrows.meta, calc, fit, backgrounds, decorations.markings, decorations.pathmorphing, patterns, shapes.misc, matrix}

\definecolor{convcolor}{HTML}{4A86E8}   
\definecolor{poolcolor}{HTML}{E06666}   
\definecolor{concatcolor}{HTML}{6AA84F} 
\definecolor{fccolor}{HTML}{1155CC}     
\definecolor{actcolor}{HTML}{F6B26B}    
\definecolor{boxblue}{HTML}{3C78D8}   
\definecolor{boxorange}{HTML}{E69138} 
\definecolor{boxpurple}{HTML}{8E7CC3}
\definecolor{convgreen}{HTML}{B5D8A3}
\definecolor{blockblue}{HTML}{9EC5E8}
\definecolor{poolgray}{HTML}{D9D9D9}
\definecolor{arrowblue}{HTML}{3B78B5}

\definecolor{matA}{RGB}{41,98,255}       
\definecolor{matB}{RGB}{230,74,25}       
\definecolor{matC}{RGB}{46,125,50}       
\definecolor{elecBg}{RGB}{232,234,246}   
\definecolor{photBg}{RGB}{255,243,224}   
\definecolor{mziGreen}{RGB}{129,199,132}
\definecolor{laserRed}{RGB}{211,47,47}
\definecolor{pdPurple}{RGB}{106,27,154}
\definecolor{dacTeal}{RGB}{0,137,123}
\definecolor{wireGray}{RGB}{120,120,120}
\definecolor{waveguide}{RGB}{255,152,0}
\definecolor{systGray}{RGB}{69,90,100}
\definecolor{compBox}{RGB}{255,235,238}  

\definecolor{digitalblue}{HTML}{D6E8F7}
\definecolor{digitalblueborder}{HTML}{4A90C4}
\definecolor{digitalbluetxt}{HTML}{1A3A5C}
\definecolor{photonicteal}{HTML}{D4F0E7}
\definecolor{photonictealborder}{HTML}{2E9E75}
\definecolor{photonictealtxt}{HTML}{0A4030}
\definecolor{pecoral}{HTML}{FAE0D4}
\definecolor{pecoralborder}{HTML}{D85A30}
\definecolor{pecoraltxt}{HTML}{6B2A10}
\definecolor{analogzone}{HTML}{FFF7ED}
\definecolor{analogborder}{HTML}{D85A30}
\definecolor{chipdash}{HTML}{888888}
\definecolor{actarrow}{HTML}{C0392B}
\definecolor{wtarrow}{HTML}{6C3483}
\definecolor{photoarrow}{HTML}{1D8348}
\definecolor{buscolor}{HTML}{2E4053}
\definecolor{meshbus}{HTML}{99A3A4}

\definecolor{inCol}{RGB}{86,180,233}
\definecolor{outCol}{RGB}{204,121,167}
\definecolor{mziCol}{RGB}{230,159,0}
\definecolor{wgColor}{RGB}{40,40,40}

\definecolor{eicfill}{HTML}{FEF2F2}
\definecolor{eicborder}{HTML}{7F1D1D}
\definecolor{eictext}{HTML}{0F172A}
\definecolor{picfill}{HTML}{DBEAFE}
\definecolor{picborder}{HTML}{1E3A8A}
\definecolor{pictext}{HTML}{0F172A}

\definecolor{pefill}{HTML}{FFFFFF}
\definecolor{peborder}{HTML}{64748B}
\definecolor{petext}{HTML}{000000}

\definecolor{analogfill}{HTML}{CBD5E1}
\definecolor{analogborder}{HTML}{334155}
\definecolor{analoglabel}{HTML}{8C6518}
\definecolor{digitalfill}{HTML}{F8FAFC}
\definecolor{digitalborder}{HTML}{94A3B8}
\definecolor{digitallabel}{HTML}{3A5478}

\definecolor{blockfill}{HTML}{FFFFFF}
\definecolor{blockborder}{HTML}{1E293B}
\definecolor{eicblockfill}{HTML}{E8EDF2}
\definecolor{eicblockborder}{HTML}{4A6785}
\definecolor{picblockfill}{HTML}{E4F2EE}
\definecolor{picblockborder}{HTML}{2E7D6A}

\definecolor{actarrow}{HTML}{1D4ED8}
\definecolor{wtarrow}{HTML}{B91C1C}
\definecolor{photoarrow}{HTML}{0F766E}
\definecolor{chipdash}{HTML}{475569}

\definecolor{chipfill}{HTML}{E6F0FA}
\definecolor{chipborder}{HTML}{004C99}
\definecolor{chiptxt}{HTML}{000000}
\definecolor{routerfill}{HTML}{FFDAB9}
\definecolor{routerborder}{HTML}{CC5500}
\definecolor{dramfill}{HTML}{F8F9FA}
\definecolor{dramborder}{HTML}{343A40}
\definecolor{buscolor}{HTML}{343A40}
\definecolor{meshbus}{HTML}{DCE1E6}
\definecolor{memborder}{HTML}{1B2631}
\definecolor{memfill}{HTML}{AEB6BF}

\definecolor{wgclad}{HTML}{D1D5DB}
\definecolor{wgcore}{HTML}{111827}
\definecolor{textcol}{HTML}{000000}
\definecolor{fadedtxt}{HTML}{6B7280}

\definecolor{lam1}{HTML}{E63946} 
\definecolor{lam2}{HTML}{D62828} 
\definecolor{lam3}{HTML}{F77F00} 
\definecolor{lam4}{HTML}{FCBF49} 
\definecolor{lam5}{HTML}{2A9D8F} 
\definecolor{lam6}{HTML}{0077B6} 
\definecolor{lam7}{HTML}{023E8A} 
\definecolor{lam8}{HTML}{7209B7} 
\definecolor{lam9}{HTML}{3A0CA3} 

\definecolor{poolgray}{RGB}{189, 215, 231}   
\definecolor{poolcolor}{RGB}{189, 215, 231}  
\definecolor{convgreen}{RGB}{107, 174, 214}  
\definecolor{convcolor}{RGB}{107, 174, 214}  
\definecolor{blockblue}{RGB}{49, 130, 189}   
\definecolor{concatcolor}{RGB}{49, 130, 189} 
\definecolor{boxblue}{RGB}{49, 130, 189}     
\definecolor{fccolor}{RGB}{253, 141, 60}     
\definecolor{actcolor}{RGB}{253, 190, 133}   
\definecolor{arrowblue}{RGB}{8, 81, 156}     
\definecolor{boxorange}{RGB}{80, 80, 80}
\definecolor{boxpurple}{RGB}{217, 217, 217}  

\usepackage[T1]{fontenc}

\usepackage{hyperref}
\hypersetup{
    colorlinks=true,
    linkcolor=blue,
    citecolor=blue,
    urlcolor=black
}

\usepackage[none]{hyphenat}

\makeatletter
\def\ps@pprintTitle{%
  \let\@oddhead\@empty
  \let\@evenhead\@empty
  \def\@oddfoot{\reset@font\hfil\thepage\hfil}
  \let\@evenfoot\@oddfoot
}
\makeatother

\begin{document}
\sloppy
\begin{frontmatter}

\title{Towards Topology-Aware Very Large-Scale Photonic AI Accelerators}

\author{Belal Jahannia}
\author{Abdolah Amirany}
\author{Hamed Dalir\corref{cor1}}
\ead{hamed.dalir@ufl.edu} 

\affiliation{organization={Department of Electrical and Computer Engineering, University of Florida},
            city={Gainesville},
            state={FL},
            country={USA}}

\begin{abstract}
The rapid growth of deep neural networks (DNNs) has exposed fundamental limitations in electronic accelerators, where data movement dominates energy consumption, commonly referred to as the memory wall. Photonic accelerators offer a compelling alternative due to their inherent parallelism and high-speed matrix operations. However, existing research largely focuses on device-level innovations, leaving system-level scalability insufficiently explored. In this paper, we present a scalable photonic accelerator architecture based on a modular scale-out paradigm using $4\times4$ photonic tensor core units. We perform a systematic architectural analysis that incorporates the practical scaling limits of photonic hardware, including insertion loss, fanout penalties, and laser power limits, which restrict monolithic photonic scaling. Through evaluation on representative DNN workloads (GoogleNet, ResNet-18, MobileNet, and AlphaGo Zero) with up to 1024 processing elements, we identify a topology-dominated scaling bottleneck in the photonic domain, termed the ``Utilization Wall'', where performance is governed by grid topology rather than hardware size. We further establish the ``Symmetric Grid Rule'', demonstrating that symmetric topologies improve utilization by up to $6\times$ while reducing memory access by over $40\%$ compared to linear configurations, which reveal that topology-aware scaling is essential for achieving energy-efficient and high-performance photonic AI accelerators.
\end{abstract}


\end{frontmatter}



\section{Introduction}
\label{sec:intro}

The rapid evolution of deep neural networks (DNNs), with model sizes reaching billions to trillions of parameters, has fundamentally transformed modern computing. However, this progress has exposed a critical limitation: the scaling boundaries of traditional electronic architectures. The emergence of specialized electronic accelerators, notably tensor processing units (TPUs)~\cite{jouppi2017datacenter, jouppi2018motivation}, has significantly improved peak computational performance for DNN workloads. Despite these advances, system efficiency is increasingly constrained by the \textit{memory wall}, where energy consumption and latency are dominated by data movement rather than computation. This phenomenon reflects a fundamental mismatch between processor throughput and memory bandwidth \cite{gholami2024aia}.

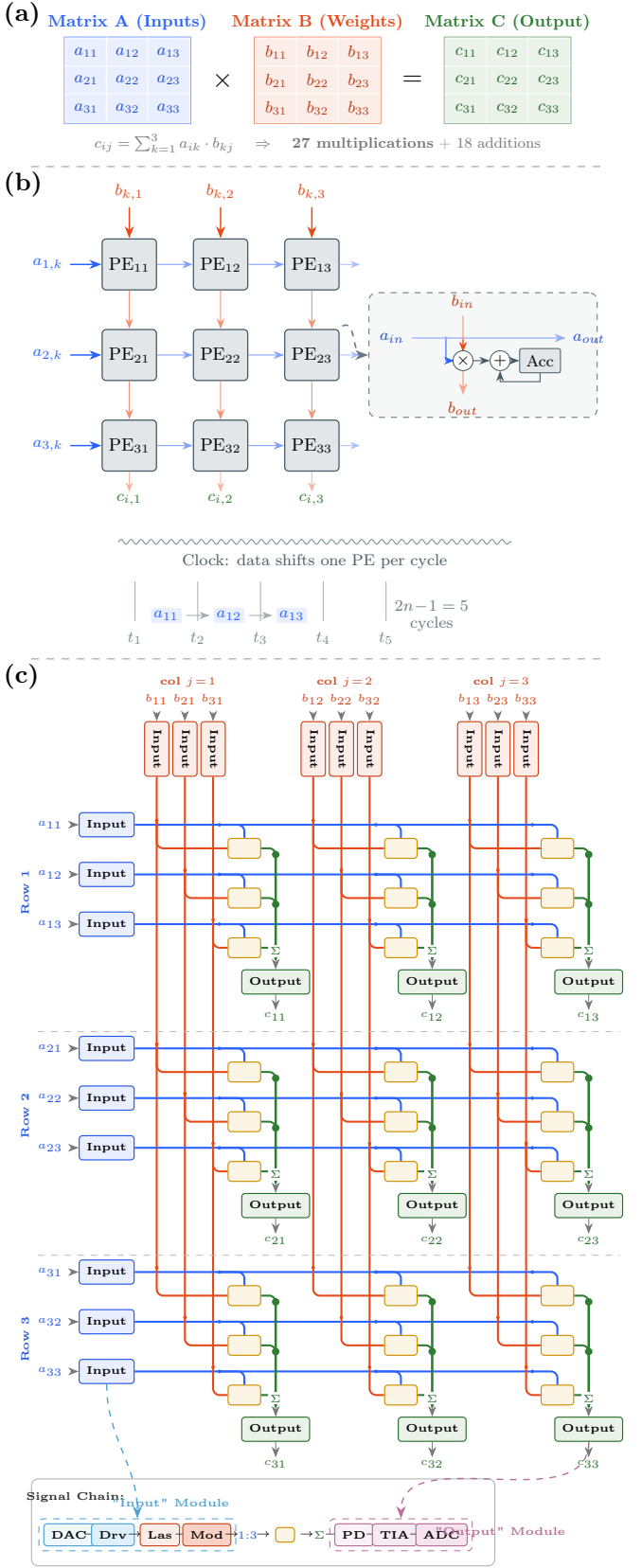
\begin{figure}[htbp] 
    \centering
    
    \makebox[\linewidth][l]{\textbf{(a)}} \\[-0.3cm]
    \resizebox{0.9\linewidth}{!}{
        \input{Figures_V03_Array/00_Matrix_V01} 
    }

    \vspace{-0.3cm}
    \noindent\tikz \draw[dashed, gray!80] (0,0) -- (0.9\linewidth,0);
    \vspace{-0.1cm}

    \makebox[\linewidth][l]{\textbf{(b)}} \\[-0.3cm]
    \resizebox{0.95\linewidth}{!}{
        \input{Figures_V03_Array/01_Electronic_V01}
    }

    \vspace{-0.3cm} 
    \noindent\tikz \draw[dashed, gray!80] (0,0) -- (0.9\linewidth,0);
    \vspace{-0.1cm}

    \makebox[\linewidth][l]{\textbf{(c)}} \\[-0.3cm]
    \resizebox{0.99\linewidth}{!}{
        \input{Figures_V03_Array/02_Photonic_V01}
    }

    \vspace{-0.2cm}
    \caption{Comparison of Matrix Multiplication Implementations: (a) Mathematical definition of a standard matrix multiplication. (b) Electronic systolic array computing partial sums sequentially ($O(n)$) clock cycles, example for 9 PEs. (c) Photonic MAC architecture executing the entire multiplication matrix simultaneously ($O(1)$) using 27 MZIs.}
    \label{fig_systolic}
\end{figure}

Integrated photonics has emerged as a promising technology to address these limitations by mitigating resistive-capacitive (RC) delays in electronic interconnects~\cite{ning2025hardware, zhou2022photonic}. Photonic accelerators can perform high-throughput matrix-vector multiplications by leveraging the inherent parallelism of optical signals~\cite{hua2025integrated,ahmed2025universal}. As illustrated in Fig.~\ref{fig_systolic}, parallel photonic multiply-and-accumulate (MAC) architectures execute matrix operations in a fully parallel manner (Fig.~\ref{fig_systolic}(c))~\cite{tait2014broadcast, zhou2022photonic}, in contrast to electronic systolic arrays that rely on sequential data propagation (Fig.~\ref{fig_systolic}(b))~\cite{jouppi2017datacenter}. By exploiting passive optical fan-out and waveguide-based accumulation, photonic systems can significantly reduce data movement overhead. Nevertheless, current photonic accelerators remain immature compared to their electronic counterparts \cite{hua2025integrated, al2022scaling}.

A critical maturity gap exists in the photonic computing literature. The majority of prior work focuses on device-level innovations, including modulators, microring resonators, and small-scale $4\times4$ photonic tensor cores (PTCs)~\cite{al2022scaling, jahannia2024low, yin2024scatter}. While these developments are essential, system-level evaluations are often limited to simplified benchmarks such as MNIST~\cite{deng2012mnist} or low-dimensional workloads. There remains a lack of systematic understanding of how to build these building blocks into scalable architectures capable of supporting complex DNNs such as GoogleNet~\cite{szegedy2015going}, ResNet-18~\cite{he2016deep}, MobileNet~\cite{howard2017mobilenets}, and AlphagoZero~\cite{silver2017mastering}.

Our goal is to bridge this gap by transitioning photonic computing from isolated device demonstrations to a system-level architectural framework. We introduce a top-down systematic architectural analysis methodology that begins from workload-level requirements and accounts for fundamental scaling limits of photonic hardware. In contrast to conventional monolithic \textit{scale-up} approaches, which are fundamentally limited by cumulative insertion loss, fanout penalties, and finite laser power budgets, we adopt a modular \textit{scale-out} strategy based on interconnected $4\times4$ PTC units.

In this paper, we propose a modular photonic scaling framework that reconciles large-scale compute requirements with physical constraints, and perform an extensive architectural exploration across four representative DNN workloads and system sizes ranging from 64 to 1024 processing elements (PEs). We show that, unlike electronic accelerators where performance scales primarily with PE count, photonic accelerators exhibit a topology-dominated scaling regime driven by optical loss and communication constraints. In this regime, system performance is governed by grid geometry rather than hardware size, leading to a scaling bottleneck specific to photonic architectures, termed the \textit{Utilization Wall}. Furthermore, we show the \textit{symmetric grid rule}, quantifying how grid symmetry and dataflow interactions minimize memory access overhead and maximize utilization.

This work focuses on the architectural interaction between grid topology, DNN workload characteristics, and memory access behavior, evaluated using a cycle-accurate simulation framework adapted from electronic systolic array analysis. To isolate these effects, certain system components such as wavelength division multiplexing (WDM), optical memory, and detailed inter-chiplet routing are abstracted, as their inclusion would require device-level modeling beyond the scope of this study. Given the cumulative loss mentioned above, each photonic chiplet is limited to a small $4\times4$ MAC array, encouraging a modular design approach. This constraint motivates our focus on understanding how such modular photonic units interact with workload structure and data movement, providing a foundation for future system-level optimizations.

The rest of the paper is organized as follows. Section~\ref{sec:background} presents background and related works. Section~\ref{sec:arch} explains our scaling philosophy and hardware building blocks. Section~\ref{sec:sims} introduces the simulation framework, workload selection, and analysis methodology. Section~\ref{sec:results} presents the main simulation results. Section~\ref{sec:discussion} discusses the key physical challenges limiting photonic scaling and outlines future research directions. Section~\ref{sec:conclusion} concludes the paper.

\section{Background}
\label{sec:background}

To achieve a scalable photonic accelerator architecture on a larger scale, a deviation from existing monolithic scaling models of electronic systolic arrays has to occur. Although electronic chips can easily scale to thousands of PEs in a monolithic chip, photonic accelerators are limited by optical signal integrity, fabrication limitations, and thermal sensitivities. This section briefly reviews prior work to support the case for modular scale-out over monolithic scaling in photonic accelerators. While system-level scalability has been well-studied in electronics, it remains an open problem in photonic computing.

Upon an analysis of existing literature in this field, a substantial gap in maturity between photonic and electronic accelerators can be observed. For instance, in the electronic domain, ``architectural scaling'' is a field that has been extensively studied and has a strong analytical basis using models such as ``SCALE-Sim V3'' \cite{raj2025scale} and industry benchmarks on ``trillion-parameter models''. However, most of the existing literature on photonic computers has been ``device-centric'', focusing on individual components and subsystems such as Mach-Zehnder Interferometer (MZI) or microring resonator (MRR) \cite{zhang2020photonic,dalir2022high_density}, modulators~\cite{wu2025highly}, photodetectors~\cite{tossoun2025large}, laser sources~\cite{cai2025tcsel}, and the supporting electronic interfaces (DAC, TIA, ADC)~\cite{amirany2025integrated}, as well as alternative material platforms such as LiNbO\textsubscript{3}~\cite{song2025integrated}.

Though these studies form a foundation, their evaluation methodologies have been restricted to basic functionality such as MNIST digit classification and small-scale data sets \cite{jahannia2024low}. However, as we demonstrate in this work, these performance improvements do not linearly extrapolate to more complex and heterogeneous data sets. Through the application of a rigorous architectural sweep to photonic computing, we bridge the gap between these two extremes. The results provide the first formal demonstration that photonic accelerators do not linearly extrapolate. Instead, they obey a set of topological constraints, dubbed the ``Symmetric Grid Rule''.

Recent work has further reinforced the physical scaling constraints that motivate our modular approach.~\cite{al2022scaling} presented a comprehensive link-budget and energy-efficiency analysis for silicon photonic accelerators. Their analysis demonstrates that cumulative optical losses impose hard upper bounds on monolithic network size, on the order of tens of elements per dimension even under optimistic assumptions. This stands in stark contrast to electronic systolic arrays, which routinely scale to thousands of PEs on a single chip. In practice, compounding insertion loss, fanout penalties, phase error accumulation, and thermal sensitivity further reduce the viable monolithic unit to the $4 \times 4$ scale adopted in this work. At the circuit level,~\cite{kim2024photonic} proposed an output-stationary photonic systolic array for all-optical matrix-matrix multiplication using adjoint designed freeform silicon modules. However, their design was numerically verified only at the $4 \times 4$ and $2 \times 2$ scales. Similarly,~\cite{tang2025waveguide} introduced a waveguide-multiplexed MVM processor using multiport photodetectors. Their fabricated $4 \times 4$ circuit achieved 93.3\% classification accuracy on the Iris dataset, a benchmark comparable in complexity to MNIST. Collectively, these studies confirm that practical photonic compute units remain confined to small monolithic dimensions, consistent with the $4 \times 4$ PTC granularity adopted in our architecture.

While these contributions address critical device- and circuit-level challenges, they do not investigate how such small-scale building blocks should be composed to execute full-scale DNN workloads. Their evaluations are limited to simple classification tasks like MNIST. Such benchmarks do not capture the heterogeneous layer structures, varying compute-to-data ratios, and complex data movement patterns characteristic of modern DNNs such as GoogleNet and ResNet-18. This gap motivates the system-level architectural analysis presented in this paper.

\section{Architectural Design}
\label{sec:arch}

This section presents the architectural building blocks that form the foundation of our scalable photonic accelerator. Based on the scaling limitations established in Section~\ref{sec:background}, we adopt a $4\times4$ PTC as the fundamental compute unit, describe the modular scale-out strategy for composing these units into larger systems, and discuss the dataflow mapping considerations that govern how data is spatially distributed across the photonic grid.

\subsection{The $4 \times 4$ Photonic Tensor Core Unit}
\label{sec:ptc4}

\begin{figure*}[!t] 
    \centering
    
    \mbox{
        \hspace{-0.05\textwidth}
        \begin{minipage}[c]{0.58\textwidth}
            \makebox[\linewidth][l]{\textbf{(a)}} \\[-0.3cm]
            \resizebox{\linewidth}{!}{
                \input{Figures/03_ScaleUp_Npe} 
            }
        \end{minipage}
        
        \hspace{0.0\textwidth}
        \begin{tikzpicture}[baseline=(current bounding box.center)]
            \draw[dashed, gray!80, line width=0.8pt] (0,-5.3) -- (0,5.2); 
        \end{tikzpicture}
        \hspace{0.00\textwidth}
        
        \begin{minipage}[c]{0.37\textwidth}
            \makebox[\linewidth][l]{\textbf{(b)}} \\[-0.3cm]
            \makebox[\linewidth][c]{
                \resizebox{0.8\linewidth}{!}{
                    \input{Figures_V02_Interposers/01_Electronic_V01}
                }
            }
            
            \vspace{-0.1cm}
            \noindent\tikz \draw[dashed, gray!80] (0,0) -- (1.1\linewidth,0);
            \vspace{-0.1cm}
            
            \makebox[\linewidth][l]{\textbf{(c)}} \\[-0.3cm]
            \resizebox{1.1\linewidth}{!}{
                \input{Figures_V02_Interposers/02_Photonics_V01}
            }
        \end{minipage}
    }

    \caption{Conceptual illustration of the Scale-Up and Scale-Out paradigms. \textbf{(a)} depicts the monolithic \textbf{Scale-Up} approach, where computing capacity is expanded by increasing the mesh dimensions within a single chiplet. While structurally simpler, this approach fundamentally exacerbates optical insertion loss and phase error accumulation. \textbf{(b)} and \textbf{(c)} depict the modular \textbf{Scale-Out} approach, where multiple chiplets, each internally structured as in (a), are tiled and interconnected to form a larger system. In (b), an electronic scale-out architecture utilizes a 2D Mesh NoC interposer for communication with off-chip memory (DRAM). In (c), a photonic scale-out architecture leverages an optical interposer for wavelength-parallel data distribution, enabling direct optical connectivity between chiplets. This contrast motivates the topology-aware scale-out methodology proposed in this work.}
    \label{fig_scale_out_interconnect}
\end{figure*}
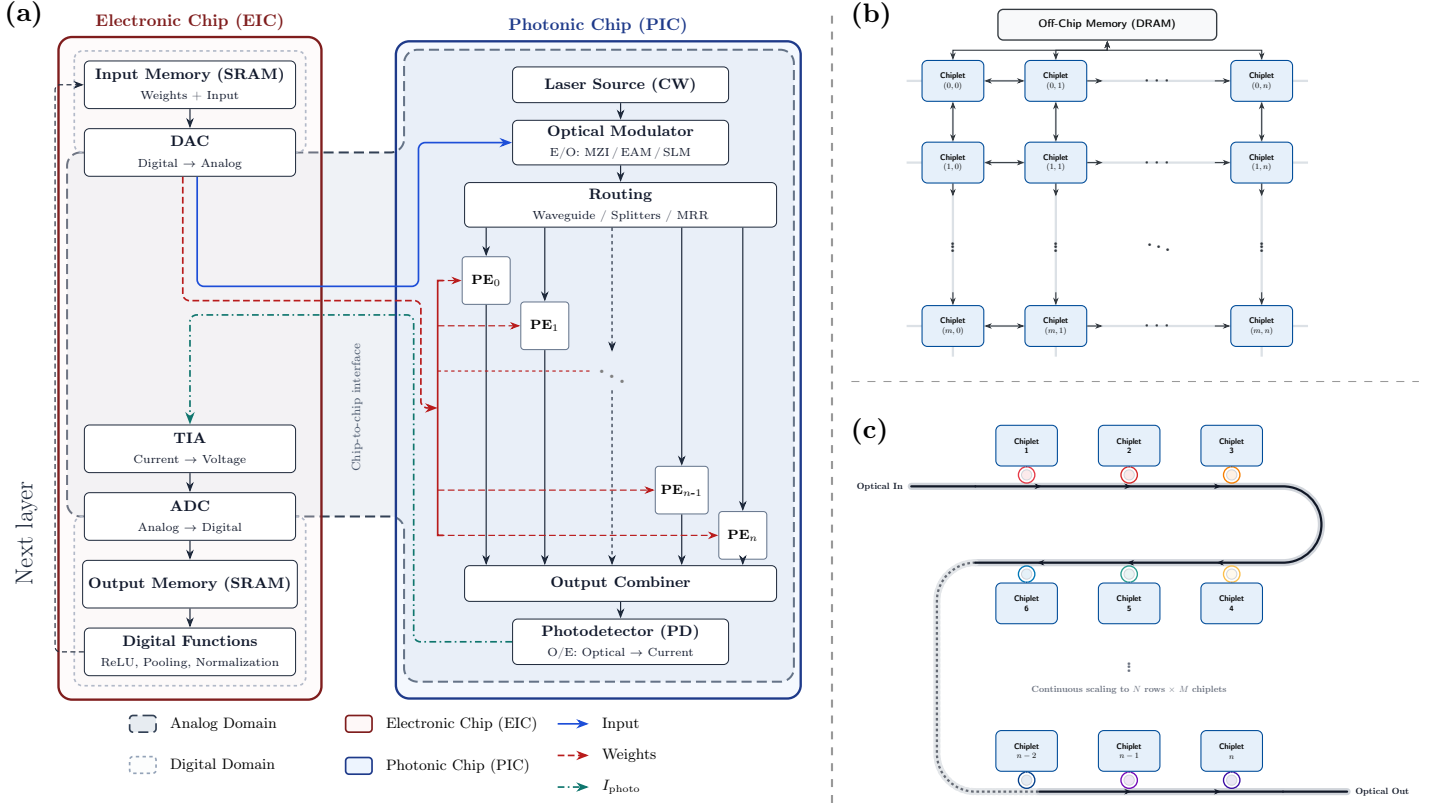

The fundamental computing element in our design is a $4\times4$ PTC that is optimized to execute high-speed matrix-vector multiplication in the photonic domain \cite{feldmann2021parallel, ma2022high}. As depicted in Fig.~\ref{fig_scale_out_interconnect}(a), each PE performs a matrix-vector MAC operation, the fundamental computation in convolutional neural networks (CNN). The physical implementation can follow either a coherent MZI mesh, where the matrix is decomposed into programmable unitary rotations, or a wavelength-parallel MRR-based architecture, where each ring weights a specific wavelength channel. The architectural analysis in this work is agnostic to the choice of PE technology and operates at the level of the $4 \times 4$ MAC tile and inter-tile grid topology \cite{ning2025hardware}.

The use of $4\times4$ granularity, rather than larger monolithic meshes such as $16 \times 16$ or $32 \times 32$, is driven by the physical limitations of silicon photonics technology. Large meshes suffer from compounding optical losses: each MZI stage contributes $0.2$--$0.5$~dB, each waveguide crossing adds $0.02$--$0.1$~dB~\cite{wu2020state}, and passive fanout of input activations to $N$ PEs costs $10 \times \log_{10}(N)$~dB ($\sim 6$~dB for $N=4$, $\sim 12$~dB for $N=16$). For a $16 \times 16$ Clements mesh, accumulated insertion loss alone reaches $\sim 6.5$~dB, which combined with fanout loss exceeds the practical $15$--$20$~dB link budget available without on-chip amplification~\cite{shafiee2023compact}. The $4 \times 4$ architecture allows for high optical signal-to-noise ratio and increases the manufacturing yield due to the reduced risk of catastrophic failures during the manufacturing phase for the larger mesh size \cite{dong2023higher}.

\subsection{Modular Scale-Out Strategy}
\label{sec:scaleout}

For obtaining large-scale computational throughput, a modular ``scale-out'' strategy is proposed in which multiple $4\times4$ PTC units are tiled and interconnected through an interposer, as shown in Fig.~\ref{fig_scale_out_interconnect}(b) and (c). Given the monolithic size constraint established in Section~\ref{sec:ptc4}, a modular scale-out strategy is adopted in which multiple $4\times4$ PTC units are tiled and interconnected through an optical interposer. By regenerating the signal at each chiplet boundary, this approach circumvents the insertion loss ceiling while maintaining optical signal integrity. Flexible grid topologies can be supported in the scale-out strategy, ranging from linear arrays ($1 \times N$) to symmetric grids ($\sqrt{N} \times \sqrt{N}$).

\subsection{Dataflow Mapping for Photonic Grids}
\label{sec:dataflow}

A central consideration in the design of any systolic accelerator is the choice of dataflow, which governs how operands are spatially and temporally mapped onto the PE array. In electronic accelerators, data stationarity, whether weights, inputs, or outputs remain resident in local storage, determines the buffering requirements and the volume of off-chip memory accesses, as extensively studied in architectures such as Eyeriss \cite{chen2016eyeriss}.

In the photonic domain, the choice of dataflow has analogous but physically distinct implications. For weight stationary (WS) mapping, the filter weights are programmed into the PTC weight elements (e.g., via phase-change materials (PCM)-programmed, MRR states or thermo optically tuned MZI phase shifters) and remain fixed for the duration of a convolutional window, while the input feature map (IFMAP) is streamed through the array. For input stationary (IS) mapping, input activations remain resident while weight matrices are reconfigured each cycle. The output stationary (OS) dataflow accumulates partial sums locally, reducing write-back traffic to off-chip memory. In the present architectural exploration, the simulation framework employs WS mapping, the dataflow strategy most naturally supported by the photonic hardware model. In this scheme, filter weights are encoded into the PCM or MZI modulators for the duration of a given convolutional window. This choice reflects a fundamental hardware asymmetry: PCM weight programming requires $10$--$100$~ns with limited endurance ($\sim 10^{6}$--$10^{8}$ cycles \cite{sawant2025high}), and thermo-optic tuning settles in $1$--$10~\mu$s, whereas input activations can be modulated at GHz rates. The architectural sweep therefore focuses on how grid topology and aspect ratio interact with this fixed-dataflow assumption. The two primary determinants of system-level efficiency, memory access overhead and PE utilization, are used as the principal evaluation metrics.

\section{Simulation Setup and Workload Selection}
\label{sec:sims}

In order to assess the scalability and performance of the proposed photonic module-based architecture, an exhaustive architectural exploration is carried out on a wide range of deep learning tasks and system configurations. Contrary to previous studies on photonic-based architectures, where only simple and artificially generated kernels were utilized, our proposed methodology focuses on realistic neural networks and their varied and complex layer stacks.

\subsection{Simulation Framework: SCALE-Sim-Photonic}
\label{sec:scalesim}

The analysis in Section~\ref{sec:results} considers how such data-mapping strategies interact with physical grid topologies to either mitigate or exacerbate the Memory Wall problem. In order to test and compare the architectures presented, a modified version of the SCALE-Sim cycle-accurate simulator is used \cite{samajdar2020systematic}. Currently, our simulation assumes an abstract memory interface. We do not explicitly model the electronic-to-photonic memory hierarchy or the specific time-of-flight limitations of preserving light states. Although the basic simulator is targeted for electronic-based systolic architectures, a modification to the basic hardware model is required to account for data transfer in a photonic system. The assumptions below separate the photonic architecture from the basic electronic systolic array simulator:

\begin{enumerate}
    \item Inter-chiplet communication is assumed to use ideal point-to-point links with a constant delay, regardless of the type of interposer technology employed.
    \item Chiplets have a limit of $4 \times 4$ MAC arrays, based on the loss budget as specified in Section~\ref{sec:ptc4}.
\end{enumerate}

This framework allows us to stress-test our modular architecture against diverse DNN workloads including GoogleNet, ResNet-18, AlphagoZero, and MobileNet, providing a rigorous assessment of scalability beyond the simple MNIST benchmarks prevalent in current photonic literature.

We quantify the efficacy of each configuration using three primary metrics derived from our modified cycle-accurate simulation framework:
\begin{itemize}
    \item \textbf{Total Compute Cycles}: In the photonic domain, a compute cycle represents one complete data transaction through the $4 \times 4$ PTC, from input modulation through optical propagation to photodetection. Since optical propagation across a millimeter-scale device occupies only a few nanoseconds, the cycle time is governed by the bandwidth of the electronic drive circuitry and the settling time of the modulation interface. The total cycle count quantifies the number of such transactions required to complete an end-to-end inference pass, including both productive MAC operations and stall cycles caused by memory access latency. Since the laser source must remain active for the entire duration, the total cycle count is the primary determinant of laser energy consumption per inference.
    \item \textbf{Average PE Utilization (\%)}: Defined as the fraction of total compute cycles during which a PTC is actively performing productive MAC operations, versus cycles during which it is stalled due to data starvation, memory fetch latency, or mapping mismatches. Unlike electronic accelerators where idle PEs simply save power, in photonic systems the laser source remains continuously powered throughout inference regardless of PE activity. Consequently, a photonic accelerator operating at 50\% utilization does not consume half the power of a fully utilized system, it operates at comparable laser power for twice as long, consuming twice the energy per inference. This makes topology-driven utilization optimization a fundamental energy-efficiency requirement, not merely a performance concern.
    \item \textbf{Memory Access Profile}: A detailed breakdown of IFMAP reads, Filter weight reads, and Output Feature Map (OFMAP) writes. This metric quantifies the memory traffic profile arising from the WS dataflow mapping employed in the simulation, enabling analysis of how grid topology modulates the volume of IFMAP reads, filter weight reads, and OFMAP writes, which constitute the three principal categories of off-chip memory traffic in the photonic domain.
\end{itemize}

By mapping these metrics across the four workloads, we provide a holistic view of how photonic hardware should be sized and shaped to meet the demands of modern artificial intelligence.

\subsection{Workload Selection}
\label{sec:workloads}

\begin{table*}[t]
\centering
\renewcommand{\arraystretch}{1.15}
\caption{Characteristics of Evaluated DNN Workloads.}
\label{tab_workload_specs}
\resizebox{\textwidth}{!}{%
\begin{tabular}{lccccccccc}
\toprule
\textbf{Network}  & \textbf{Family} & \textbf{Domain} & \textbf{Params (M)} & \textbf{FLOPs (B)} & \textbf{Top-1 (\%)} & \textbf{Size (MB)} & \textbf{Layers} & \textbf{Kernel Types} & \textbf{Layer Homogeneity} \\
\midrule
GoogleNet   & CNN          & Image Classif.    & 6.8      & 1.50 & 69.8 & 27.2     & 22  & 4 & Very Low  \\
ResNet-18   & CNN          & Image Classif.    & 11.7     & 1.80 & 69.8 & 46.8     & 18  & 2 & High      \\
MobileNet   & CNN (Effic.) & Mobile / Edge     & 4.2      & 0.57 & 70.6 & 16.8     & 28  & 2 & Medium    \\
AlphaGoZero  & Deep RL      & Game  & 10.0 & 7.5  & N/A  & 40.00 & 40 & 2 & Very High \\
\bottomrule
\end{tabular}
}%
\end{table*}

We have selected four representative DNN models that exhibit diverse architectural characteristics, including model depth, branching, and parameter density. This selection is crucial because we want to understand the interplay among diverse connectivity patterns with the data movement mechanism based on the photonic accelerator grid. The workloads are summarized in Table~\ref{tab_workload_specs}:

\begin{enumerate}
    \item \textbf{GoogleNet \cite{szegedy2015going}}: Characterized by its inception modules, this network features significant branching and varying kernel sizes ($1 \times 1$, $4\times4$, $5 \times 5$), providing a rigorous test for grid utilization during non-uniform layer execution.
    \item \textbf{ResNet-18 \cite{he2016deep}}: A standard residual network consisting of sequential convolutional blocks and skip connections. It serves as a baseline for deep, heavy-compute architectures with uniform kernel distributions.
    \item \textbf{MobileNet \cite{howard2017mobilenets}}: Designed for mobile efficiency, this model utilizes depthwise separable convolutions. It represents models with high memory-to-compute ratios, which are particularly sensitive to the ``Memory Wall'' in photonic interconnects.
    \item \textbf{AlphaGoZero \cite{silver2017mastering}}: A reinforcement learning model featuring deep residual towers. This workload is utilized to evaluate scaling performance on extremely deep, repetitive architectural patterns.
\end{enumerate}

\subsection{Systematic Architectural Analysis}
\label{sec:sweep}

The scalability analysis process entails sweeping two major dimensions: the total number of PEs and the grid aspect ratio. For the PE scaling dimension, we consider system sizes of 64, 128, 256, 512, and 1024 PEs, where each PE acts as a $4 \times 4$ PTC block as described in Section~\ref{sec:background}, enabling us to characterize how PE utilization and memory access overhead evolve at higher PE counts. For the topology dimension, we explore various grid aspect ratios at each PE count to study the effect of grid geometry on data reuse and memory access behavior. For instance, at 128 PEs, configurations range from highly asymmetric linear arrays such as $1 \times 128$ to near-square grids such as $8 \times 16$ and $4 \times 32$. As described in Section~\ref{sec:results}, this sweep is central to discovering the ``Symmetric Grid Optimality'' principle that balances interposer communication bandwidth with local memory read rates.

\begin{figure*}[!t]
    \centering
    \subfloat[\label{fig:googlenet_arch}]{%
        \begin{tikzpicture}[remember picture]
            \node[inner sep=0pt] (archfig) {%
                \resizebox{0.9\textwidth}{!}{%
                    \rotatebox{270}{%
                        \input{Figures/01_Fig__GoogLeNet_V00}%
                    }%
                }%
            };
            \node[coordinate] (target_a) at ([xshift=1.8cm, yshift=-2.85cm]archfig.center) {};
        \end{tikzpicture}%
    }

    \vspace{-1.2em}
    \subfloat[\label{fig:googlenet_perf}]{%
        \begin{tikzpicture}[remember picture]
            \node[inner sep=0pt] (perffig) {%
                \includegraphics[width=\textwidth]{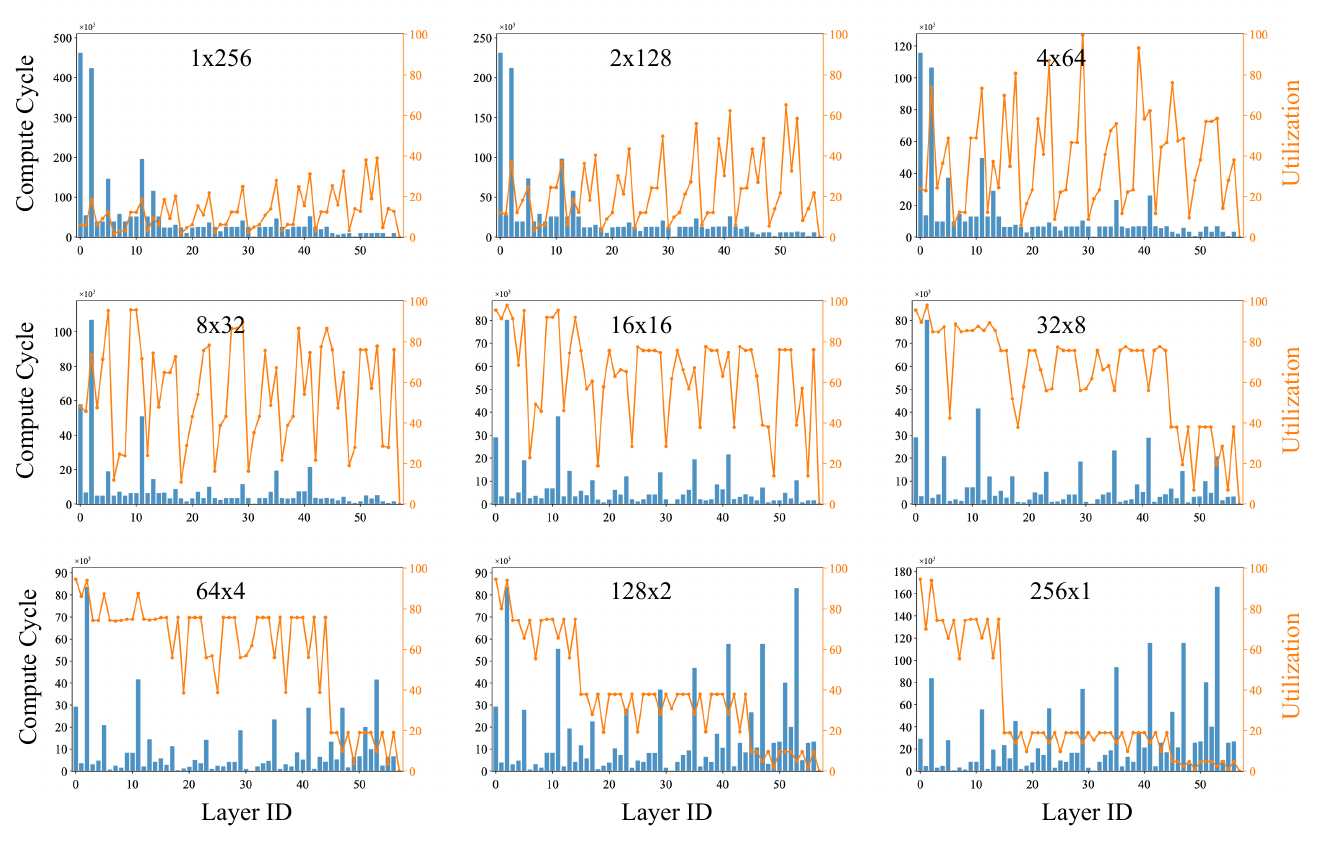}%
            };
            \node[draw=red, thick, rounded corners=2pt, 
                  minimum width=0.55cm, minimum height=10.5cm] 
                  (box_b) at ([xshift=0.50cm, yshift=-0.0cm]perffig.center) {};
        \end{tikzpicture}%
    }
    \begin{tikzpicture}[remember picture, overlay]
        \draw[-{Stealth[length=3mm]}, red, thin] 
            (box_b.north) -- (target_a.south);
    \end{tikzpicture}
    \caption{GoogLeNet system analysis. (a) Displays the detailed mapping of the GoogLeNet architecture, highlighting the Inception modules and auxiliary branches, while (b) shows layer-wise performance metrics for 256 PEs grid configurations.}
    \label{fig_googlenet_combined}
\end{figure*}

\section{Results and Architectural Analysis}
\label{sec:results}

Herein, we provide an extensive evaluation of the architectural scaling behavior of modular photonic accelerators, focusing exclusively on the interaction between grid topology, DNN workload characteristics, the resulting memory access patterns, and PE utilization. This evaluation will commence with an intrinsic level-wise evaluation of the performance to understand the fundamental relationship between the depth of the NN and the hardware's topology. Then, we will expand our evaluation to diverse PE sizes and configuration aspect ratios to understand the physical laws that govern photonic scalability.

\subsection{Layer-wise Performance Analysis}
\label{sec:layerwise}

The architectural efficiency of a photonic accelerator is mostly determined by how well the physical grid topology matches the characteristics of the data movement demands of particular neural network layers. In order to characterize the empirical scaling behavior of the photonic accelerator, specifically how layer-by-layer computational efficiency varies with grid topology across heterogeneous DNN stages, granular layer-wise execution profiles are extracted from the simulation data and analyzed across the full depth of each benchmark network.

Our simulation findings indicate a profound sensitivity to grid aspect ratios across all evaluated workloads. Near-square configurations consistently achieve the highest PE utilization, whereas even modest departures from symmetry cause disproportionate performance degradation. For instance, transitioning from a $16 \times 16$ to a $32 \times 8$ configuration already incurs a measurable utilization penalty, and the effect amplifies sharply as configurations approach the fully linear extremes ($1 \times N$ or $N \times 1$). For linear topologies, layer-averaged PE utilization collapses to below 15\% even when the aggregate PE count is held constant. This demonstrates that maximizing the total number of PEs, without matching the topological arrangement to the workload's data reuse pattern, fails to translate additional hardware into proportional performance gains for photonic accelerators.

To illustrate how this topological sensitivity manifests across different network architectures, we examine the layer-wise behavior of GoogleNet and ResNet-18 on a fixed 256-PE grid across a range of configurations.

\textbf{GoogleNet (Heterogeneity, Mapping Stalls):} The heterogeneous nature of GoogleNet, composed of Inception blocks, creates a very heterogeneous layerwise workload. As shown in Fig.~\ref{fig_googlenet_combined}, the $1\times256$ linear configuration exhibits severe utilization catastrophes. In particular, during the initial convolutional stages, the utilization is less than $15\%$, while compute cycles are over $4 \times 10^5$ for these stages. As shown through our results, these compute cycle spikes are not due to computational complexity, but rather the inability to efficiently reuse memory within a linear configuration for branchy models. In comparison, the $16\times16$ symmetric grid configuration exhibits a stable utilization floor, completing these same layers in $\leq 8 \times 10^3$ compute cycles, a nearly $5\times$ improvement.

This behavior traces directly to the Inception modules: the parallel $1 \times 1$, $3 \times 3$, and $5 \times 5$ branches produce layers with widely varying compute-to-data ratios. The $1 \times 1$ convolutions generate minimal computation per input element and cannot fill a linear array, leaving most PEs idle. The symmetric grid mitigates this by enabling two-dimensional data reuse across both spatial and channel dimensions, dampening the utilization fluctuations visible in Fig.~\ref{fig_googlenet_combined}(b).

\begin{figure*}[!t]
    \centering

    \subfloat[\label{fig:resnet_arch}]{%
        \begin{tikzpicture}[remember picture]
            \node[inner sep=0pt] (archfig) {%
                \resizebox{0.9\textwidth}{!}{%
                    \rotatebox{0}{%
                        \input{Figures/02_ResnetFig}%
                    }%
                }%
            };
            \node[coordinate] (target_a) at ([xshift=1.8cm, yshift=-1.9cm]archfig.center) {};
        \end{tikzpicture}%
    }
    
    \vspace{-1.2em}

    \subfloat[\label{fig:resnet_perf}]{%
        \begin{tikzpicture}[remember picture]
            \node[inner sep=0pt] (perffig) {%
                \makebox[0.9\linewidth][c]{%
                    \includegraphics[width=\linewidth]{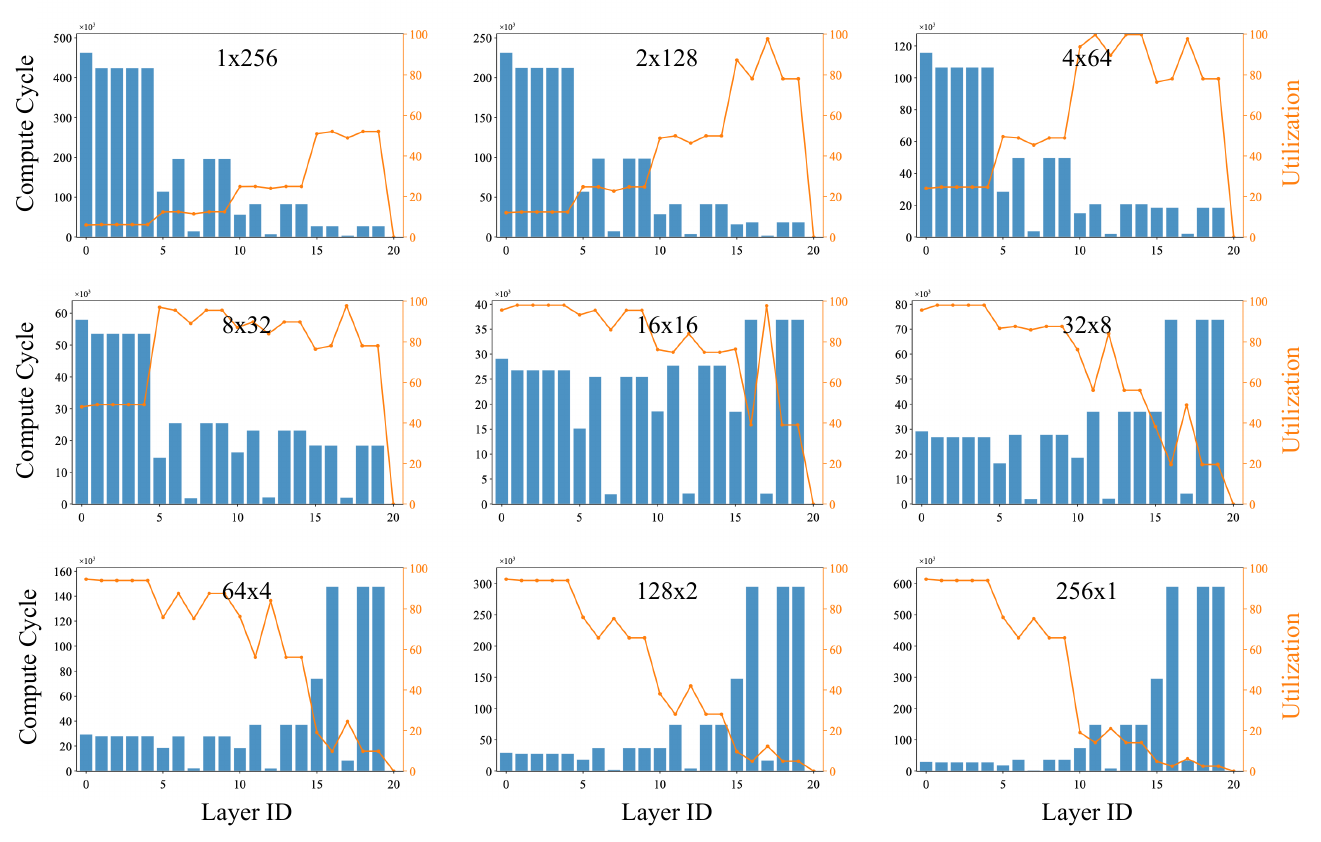}
                }%
            };
            \node[draw=red, thick, rounded corners=2pt, 
                  minimum width=1.3cm, minimum height=10.6cm] 
                  (box_b) at ([xshift=1.42cm, yshift=0.1cm]perffig.center) {};
        \end{tikzpicture}%
    }

    \begin{tikzpicture}[remember picture, overlay]
        \draw[-{Stealth[length=3mm]}, red, thin] 
            (box_b.north) -- (target_a.south);
    \end{tikzpicture}

    \caption{ResNet-18 system analysis. (a) Displays the detailed flow of the ResNet-18 architecture network, while (b) shows the layer-wise performance on a 256-PE grid. Asymmetric topologies consistently underperform, highlighting the necessity of symmetric grid layouts for sequential convolutional blocks.}
    \label{fig_resnet_256}
\end{figure*}
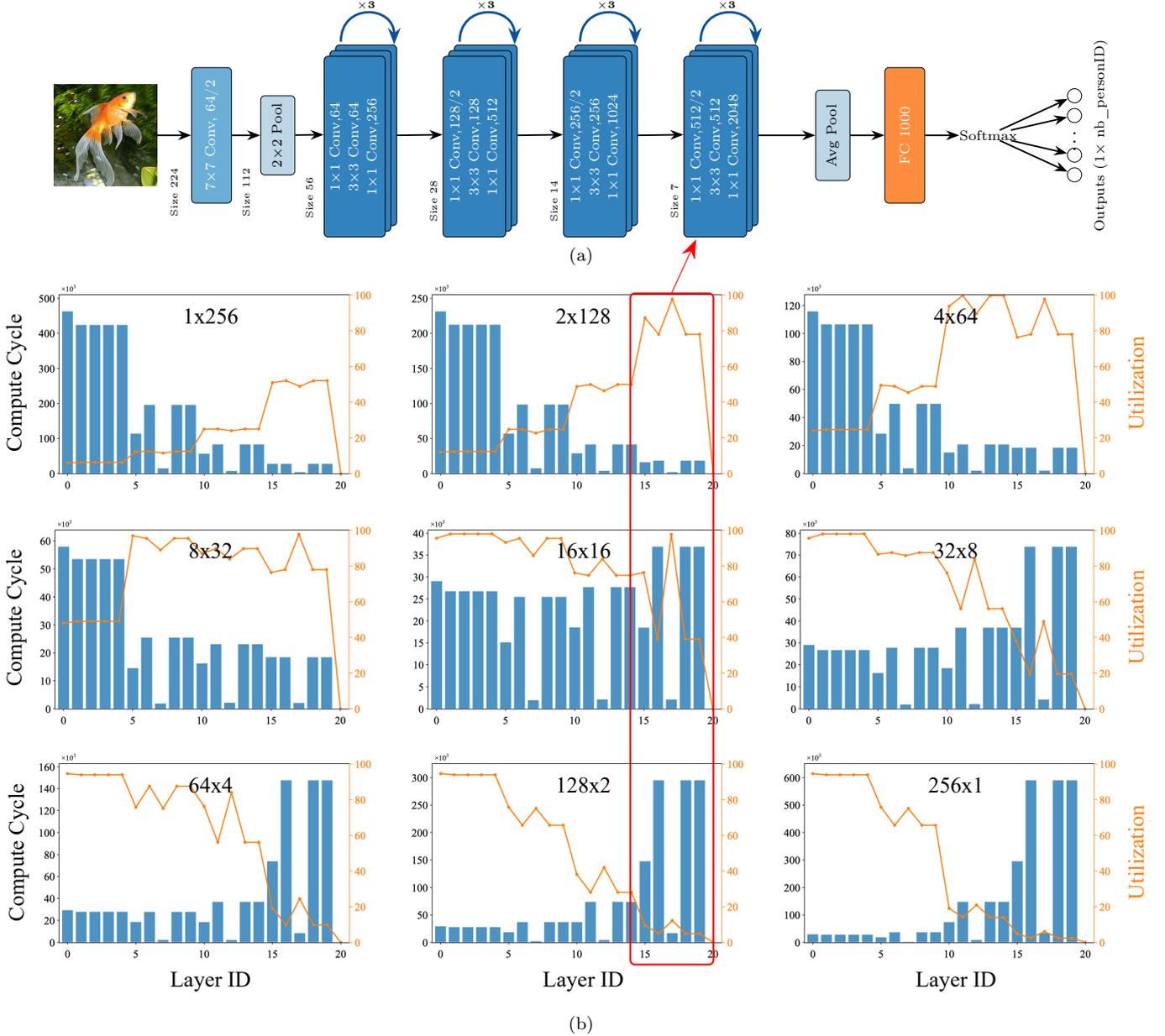

\textbf{ResNet-18 (Uniformity and Symmetry Benefits):} Although the uniformity of the performance characteristics is better with ResNet-18 due to the repetition of the residual blocks, the topological arrangement is as critical as that in the previous case. From the performance characteristics depicted in Fig.~\ref{fig_resnet_256}, it is clear that the $1\times256$ and $2\times128$ topologies are the bottlenecks across the full depth of the network, consistently providing the maximum cycle counts and minimum utilization rates. The $16\times16$ square topology is consistently the Pareto optimal solution, maintaining high utilization rates while minimizing the time taken to execute each layer.

The topology sensitivity in ResNet-18 arises from the progressive reduction in spatial dimensions ($56 \times 56 \rightarrow 7 \times 7$) paired with increasing channel depth ($64 \rightarrow 512$). As seen in Fig.~\ref{fig_resnet_256}(b), early layers with large feature maps dominate cycle counts in column-heavy topologies ($1 \times 256$), while later layers with small spatial dimensions ($7 \times 7$ = 49 elements) cannot fill a row-heavy array ($256 \times 1$), causing the utilization collapse in the final layers. The $16\times16$ topology balances both regimes.

From the performance characteristics, it is also clear that spatial symmetry of the grid is critical for achieving an efficient match between the interposer communication bandwidth and the processing capability of the individual $4\times4$ PTC blocks, particularly for sequential CNNs.

\subsection{System-Level Scaling and Topological Sensitivity}
\label{sec:scaling}

\begin{figure*}[!t]
    \centering
    \includegraphics[width=0.95\textwidth]{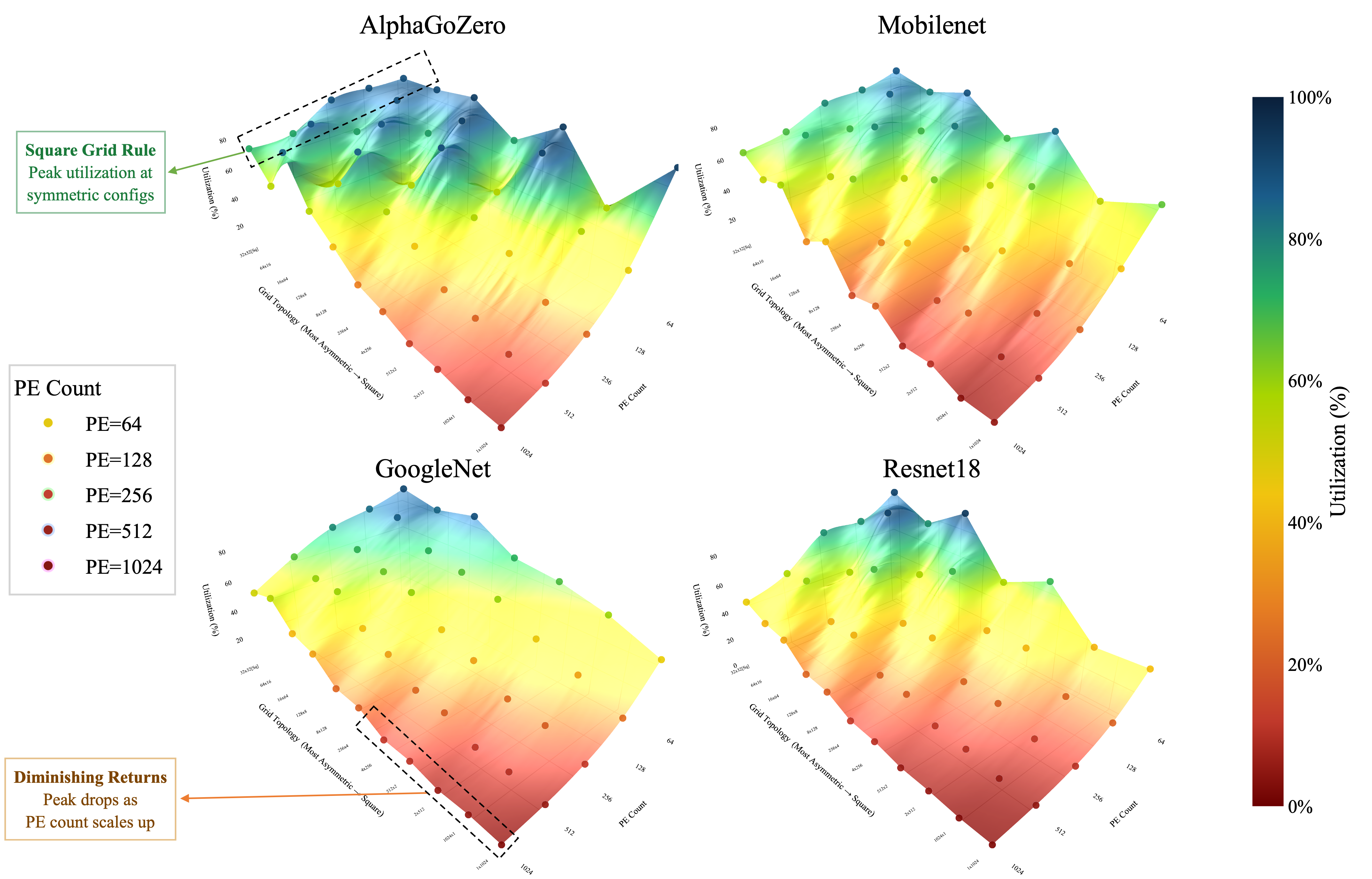}
    \caption{System-level PE utilization (\%) across four benchmark DNNs as a function of grid configuration (x-axis) and PE count (depth axis). Each subplot represents a specific workload. Peak utilization consistently occurs at symmetric (square or near-square) grid configurations. The `Diminishing Returns' annotation highlights the progressive utilization decline as PE count increases beyond 512.}
    \label{fig_all_pe_results}
\end{figure*}

In order to assess the macroscopic level performance of the modular photonic architecture, a detailed design space sweep was performed over a range of five PE sizes, including 64, 128, 256, 512, and 1024, as well as a range of grid aspect ratios. To quantify scaling efficiency, we define $\eta = \frac{T_{\text{base}} / T_{\text{scaled}}}{N_{\text{scaled}} / N_{\text{base}}}$, where $T$ is total execution cycles and $N$ is the PE count; $\eta = 1$ indicates ideal linear scaling. Figure~\ref{fig_all_pe_results} depicts the results, where the average PE utilization is shown for each workload, grouped by dataset, along with the scale of the overall system. This reveals the essential scaling laws that are critical to large-scale photonic computing.

A noticeable trend that emerges with all datasets is the extreme sensitivity of the performance with regard to the grid structure, regardless of the overall PE count. Highly asymmetrical configurations, i.e., $1 \times N$ or $N \times 1$ linear strings, always represent the worst cases. For example, in the AlphaGoZero workload on the 512-PE system, the optimal $16 \times 32$ configuration completes execution in $4.93 \times 10^4$ cycles and achieves 87.15\% peak utilization. By contrast, the linear configuration of $512 \times 1$ results in the utilization crashing to just $14.05\%$ while the execution cycles blow up six-fold to $3.06 \times10^5$. Similarly, in the context of the ResNet-18 workload on the same scaled system, the linear configuration of $512 \times 1$ results in the execution of the workload while consuming more than $2.84 \times10^6$ cycles, while the optimal configuration of $16 \times 32$ results in the execution of the workload in just $3.00 \times10^5$ cycles.

Beyond topology, the scale of the system itself introduces diminishing returns. Although increasing the number of PE from 64 to 1024 results in a reduction in total compute cycles, there are significant diminishing returns associated with such improvement. Our simulation results show that as the number of modular $4\times4$ blocks increases, the ability of the mapping logic to sustain high utilization decreases. In the GoogleNet workload, scaling from 128 PEs ($8 \times 16$) to 1024 PEs ($32 \times 32$) reduces total execution cycles from $7.73 \times 10^5$ to $1.59 \times 10^5$. However, this $8\times$ increase in hardware resources yields only a $4.8\times$ speedup, $(\eta = 0.60)$, while peak utilization drops from $85.29\%$ to $51.67\%$. By comparison, AlphaGoZero maintains $\eta = 0.78$ over the same range due to its uniform residual structure, while MobileNet degrades to $\eta = 0.42$. The peak utilization simultaneously drops from 85.29\% to 51.67\%, illustrating the onset of diminishing returns. This trend suggests that beyond the 512-PE scale, the workload partitioning overhead and interposer bandwidth constraints begin to outweigh the raw computational gains of additional PEs.

The robustness of the scaling strategy differs for various model architectures. As shown in Fig.~\ref{fig_all_pe_results}, scaling robustness varies significantly across workloads and correlates directly with layer homogeneity. AlphaGoZero's 40 repetitive residual blocks (all $3 \times 3$ convolutions with uniform channel depth) produce consistent compute-to-data ratios across layers, enabling near-square topologies to maintain high utilization even at 1024 PEs. GoogleNet, by contrast, shows a steady utilization decline with increasing PE count due to its Inception modules, the mix of $1 \times 1$, $3 \times 3$, and $5 \times 5$ branches creates layers that vary widely in their ability to fill the PE array, and this mismatch worsens as more PEs are added. MobileNet's depthwise separable convolutions produce extremely low compute-to-data ratios, limiting peak efficiency to approximately 72.85\% at 512 PEs ($16 \times 32$) and degrading sharply beyond this point. These results indicate that the optimal PE count and topology are workload-dependent, emphasizing the need for reconfigurable photonic interconnects.

\begin{figure*}[!t]
    \centering
    \includegraphics[width=\textwidth]{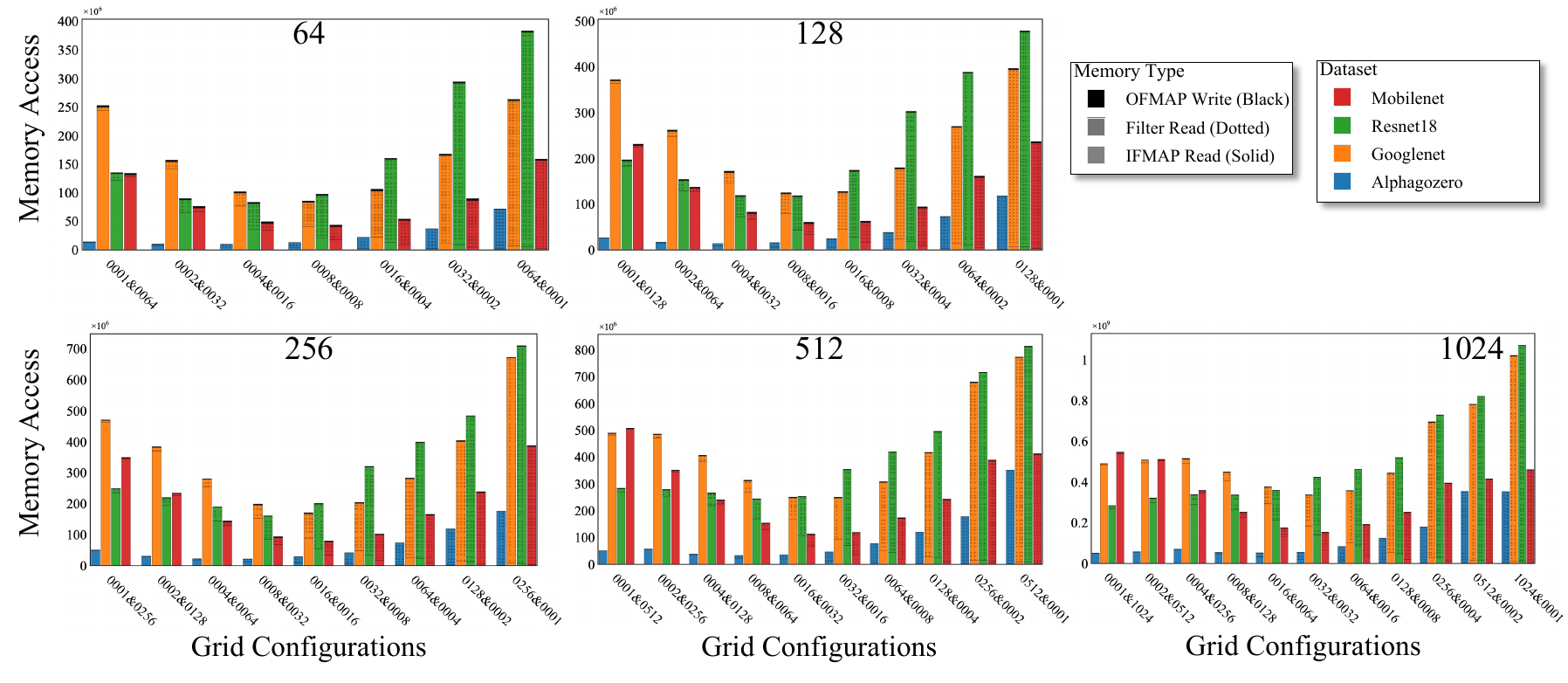}
    \caption{Breakdown of memory access patterns (IFMAP reads, Filter reads, and OFMAP writes) across different PE scales and grid configurations. The results quantify the sensitivity of off-chip memory traffic to grid topology, illustrating the architectural dimension of the Memory Wall problem.}
    \label{fig_mem_access_all}
\end{figure*}

\subsection{Memory Access Patterns}
\label{sec:memory}

The major bottleneck in the current DNN accelerator architectures is the energy and latency costs associated with the external memory accesses. In photonic architectures, the Memory Wall is a particularly acute concern. Every read operation from external memory requires an E/O conversion at the modulator, a process that is a major contributor to the overall system power budget. In the following section, the grid topology and scaling are discussed with regard to the overall data movement operations involving the IFMAP and the filter weights.

\textbf{Dataflow-Topology Interaction:} As can be inferred from the memory access profile results on the four datasets, as shown in Fig.~\ref{fig_mem_access_all}, a striking correlation is observed between grid symmetry and memory fetch efficiency. An important observation is that very asymmetric grid structures cause an explosion in memory fetches. This is because of the limited opportunity for data reusing in linear structures. An important architectural finding is that the relationship between grid aspect ratio and memory fetch overhead is strongly non-linear; for example, at the 512-PE scale, filter reads increase by up to $18\times$ between the symmetric $16\times32$ and fully linear $512\times1$ configurations. When we consider grid topologies that are increasingly asymmetric relative to their symmetric and square configurations (e.g., $16\times16$ versus $1\times256$ or $256\times1$), we see an escalation of redundant IFMAP and filter weight reads. In the layers where the data-per-computation ratio is high, the asymmetric arrangements achieve an increase of up to $4\times$ in the overall memory access latency as compared to their square counterparts. In the photonic domain, where off-chip data transfers require costly E/O and optical-to-electrical (O/E) conversions at chiplet boundaries, memory stalls directly translate to wasted computation cycles and increased energy consumption.

As an example of this effect, in the AlphaGoZero workload running at the 512-PE level of granularity, the symmetric configuration of the grid as a $16 \times 32$ grid incurs $1.58 \times 10^7$ IFMAP reads and $1.94 \times 10^7$ filter weight reads. However, as the grid is ``stretched'' in the asymmetric configuration of $512 \times 1$, the filter weight reads skyrocket to $3.49 \times 10^8$, an increase of nearly $18 \times$. This ``asymmetry penalty'' demonstrates the key point of the work: regardless of the optical interconnect mechanism employed, the physical grid geometry fundamentally controls whether data can be efficiently reused across the PEs or whether it must be repeatedly re-fetched from off-chip memory.

\textbf{The Symmetric Grid Advantage:} As observed with the memory access patterns of MobileNet, as depicted in Fig.~\ref{fig_mem_access_all} for MobileNet, square or near-square configurations have consistently provided the necessary balance of bandwidth for the input and the weights. For the 512-PE system for MobileNet, the $32 \times 16$ configuration provides $3.35 \times 10^7$ IFMAP reads and $8.20 \times 10^7$ filter reads. If the configuration were changed to $512 \times 1$, the number of reads for the filters becomes $4.06 \times 10^8$, creating a memory-side bottleneck for the compute units. This balancing act is essential for photonic scalability. The optical interposer operates under finite communication bandwidth, constrained by both the available laser power and the insertion loss budget. An explosion in data fetch volume, as observed in the linear arrangements, saturates the interposer bandwidth and prevents effective overlap of the computation phase and the communication phase. Symmetric grids minimize the amount of data, thus reducing the stress on the E/O modulators, which results in a drastic reduction in the energy per inference.

\textbf{Scaling Impact on Memory Overhead:} Furthermore, as can be observed from Fig.~\ref{fig_mem_access_all}, as the total PE count increases from 64 to 1024, the relative effect of ``Symmetric Rule'' increases in proportion. At the 64-PE scale, the difference in memory access between optimal and suboptimal configurations is modest and does not critically affect power consumption. At the 1024-PE scale, however, a suboptimal configuration can produce memory access volumes that are orders of magnitude larger than necessary for the same computational workload. For instance, in terms of memory access required to perform a ResNet-18 workload, a change from a square grid of size $32 \times 32$ to a linear grid of size $1024 \times 1$ increases filter reads from $2.82 \times 10^8$ to over $1.05 \times 10^9$. These results highlight that in a larger grid size of a photonic accelerator system, architecture is not only a performance booster but a necessity to operate below a system power constraint.

\subsection{Whole System Performance Analysis}
\label{sec:wholesystem}

\begin{table}[t]
\centering
\caption{Comparison of Digital, Analog, and Photonic Architectures Across Workloads.}
\label{tab:arch_compare}
\setlength{\tabcolsep}{4pt}
\renewcommand{\arraystretch}{1.3}
\footnotesize
\begin{tabular}{@{}cl cccc@{}}
\toprule
\textbf{Arch.} & \textbf{Metric}
  & \textbf{AlphaGo} & \textbf{Mobile} & \textbf{GoogLe} & \textbf{Res} \\
 & & \textbf{Zero} & \textbf{Net} & \textbf{Net} & \textbf{Net} \\
\midrule
\multirow{4}{*}{\rotatebox[origin=c]{90}{\scriptsize Digital~\cite{tu202228nm}}}
  & Energy (fJ/op)             & \multicolumn{4}{c}{\textbf{34.2}} \\
  & TOPS/W                     & \multicolumn{4}{c}{\textbf{29.2}} \\
  & Eff.\ Thr.\ ($\propto\!f$) & 6.3$f$ & 5.4$f$ & 4.8$f$ & 4.4$f$ \\
  & Eff.\ TOPS/W               & 17.8 & 15.4 & 13.6 & 12.5 \\
\midrule
\multirow{4}{*}{\rotatebox[origin=c]{90}{\scriptsize Analog~\cite{tu202228nm}}}
  & Energy (fJ/op)             & \multicolumn{4}{c}{149.2} \\
  & TOPS/W                     & \multicolumn{4}{c}{6.7} \\
  & Eff.\ Thr.\ ($\propto\!f$) & 8.9$f$ & 7.5$f$ & 6.8$f$ & 6.1$f$ \\
  & Eff.\ TOPS/W               & 5.8 & 4.9 & 4.5 & 4.0 \\
\midrule
\multirow{4}{*}{\rotatebox[origin=c]{90}{\scriptsize Photonic}}
  & Energy (fJ/op)             & \multicolumn{4}{c}{47.2} \\
  & TOPS/W                     & \multicolumn{4}{c}{21.2} \\
  & Eff.\ Thr.\ ($\propto\!f$) & \textbf{71.4}$f$ & \textbf{59.7}$f$ & \textbf{54.6}$f$ & \textbf{48.9}$f$ \\
  & Eff.\ TOPS/W               & \textbf{18.5} & \textbf{15.4} & \textbf{14.1} & \textbf{12.7} \\
\bottomrule
\end{tabular}
\end{table}

Digital and analog accelerators differ fundamentally in how multiply operations are implemented and where energy is consumed. Digital CMOS implementations rely on Boolean switching in arithmetic units and have achieved high efficiency through technology scaling and optimized datapaths, with reported efficiencies reaching tens of TOPS/W for low-precision inference~\cite{tu202228nm}. In contrast, analog and mixed-signal accelerators perform computation in the charge or memory domain, significantly reducing core computation energy. However, their overall system efficiency is often limited by the overhead of data conversion, particularly analog-to-digital (ADC) and digital-to-analog (DAC) converters, which dominate power consumption in many practical implementations~\cite{le202364}.

For a fair comparison across different computing paradigms, digital and analog architectures are implemented using a fixed $32\times32$ array configuration. Prior work has shown that square and balanced array organizations provide near-optimal performance for systolic and matrix-based accelerators by minimizing boundary inefficiencies and maximizing dataflow regularity~\cite{yuzuguler2023scale}. Therefore, adopting a $32\times32$ architecture ensures a representative and efficient baseline for digital and analog implementations, enabling a meaningful comparison with the proposed photonic architecture.

The results in Table~\ref{tab:arch_compare} reflect these architectural differences. Digital implementations achieve the lowest energy per operation due to highly optimized arithmetic units and mature design flows~\cite{tu202228nm}. Analog implementations exhibit higher energy per operation at the system level, despite efficient in-memory computation, due to the significant contribution of ADC/DAC overheads~\cite{le202364}. These trends are consistent across workloads and align with previously reported system-level measurements in state-of-the-art digital and mixed-signal accelerators.

Compared to these electronic implementations, photonic architectures achieve substantially higher effective throughput due to inherent parallelism, while maintaining competitive energy efficiency. Although photonic systems incur additional interface overhead, their ability to perform large numbers of operations simultaneously enables superior throughput-driven performance, making them particularly advantageous for large-scale AI workloads.

\section{Challenges and Future Directions}
\label{sec:discussion}

The move from isolated device-level demonstrations toward a unified system-level architecture signifies a turning point in the road toward photonic computing. Within the following section, we place the results within the context of the overall hardware acceleration paradigm, tackle the current technological roadblocks as determined by this work, and offer a roadmap toward the creation of a photonic tensor processor.

\begin{figure*}[!t]
    \centering
    \includegraphics[width=0.9\textwidth]{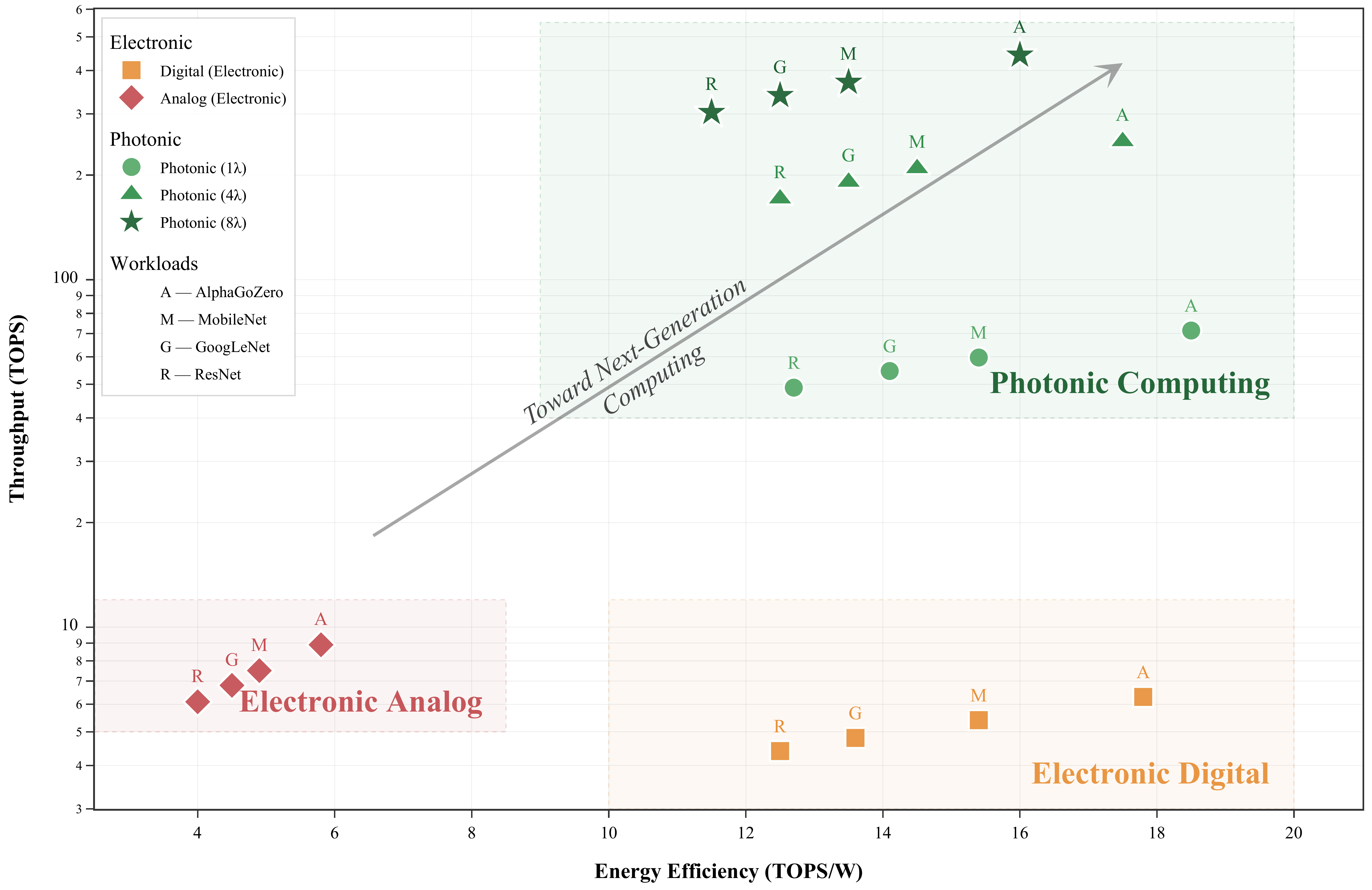}
    \caption{Throughput versus energy efficiency for electronic and photonic accelerators across four DNN workloads. Wavelength-multiplexed photonic operation (1$\lambda → 4\lambda → 8\lambda$) extends the throughput frontier beyond electronic digital and analog regimes.}
    \label{fig_pe_layout}
\end{figure*}

\subsection{The Interconnect and Switching Bottleneck}
\label{sec:interconnect}

While photonic switch technologies exist, including thermo-optic MZI switches ($\sim 10$--$100~\mu$s switching time), electro-optic switches ($\sim 1$--$10$~ns), and micro-electromechanical (MEMS)-based crossbars ($\sim 1$--$10$~ms) \cite{stepanovsky2019comparative}, these have been developed for telecommunications routing, not for the high-rate, per-layer reconfiguration patterns required by DNN inference workloads. In an established electronics-based architecture like Eyeriss, highly sophisticated spatial switching and buffering are implemented to optimize data reuse for IS and WS mapping \cite{chen2016eyeriss}. In a photonic architecture such as the one explored in this work, the optical interposer provides inter-chiplet connectivity; however, efficient data reuse and broadcasting remain major challenges.

As shown in our evaluation in Section~\ref{sec:results}, highly asymmetric grid configurations incur significant stall penalties because the interposer communication capacity cannot efficiently sustain the data distribution demands of the PEs. To transcend these ``logical'' mapping limitations, future research must focus on building lossless photonic crossconnects, such as non-blocking Benes topologies. These switching fabrics must be designed to handle the high data rates of neural network inference workloads, rather than the lower-rate patterns typical of traditional telecommunications applications \cite{jahannia2024low, zhang2020photonic}. A key open question is the choice of interposer technology: a photonic interposer keeps data in the optical domain between chiplets but requires signal regeneration to compensate accumulated loss, while an electronic interposer requires O/E/O conversion at each chiplet boundary but benefits from high SNR and mature fabrication processes.

\subsection{Physical Constraints: The Thermal and Loss Walls}
\label{sec:physical}

While our results demonstrate the presence of throughput maxima in the theory of 1024-PE configurations, the feasibility of these levels of scaling is challenged from the perspective of thermal as well as optical limits. There are two principal physical constraints limiting the scaling up of photonic accelerators as explained below.

This level of density in the PE grid also leads to large localized heat signatures. In silicon photonics, the refractive index is very sensitive to temperature, and this leads to a wavelength shift with every 1$^\circ$C rise in the system, or 0.1~nm per degree Celsius~\cite{parra2024silicon}. In a dense WDM system, a 5$^\circ$C temperature difference across the interposer leads to crosstalk between different wavelengths, ruining the matrix multiplication. In fact, our results in Section~\ref{sec:scaling}, which show that utilization fails for PE counts over 512, indicate that such systems are not only inefficient but perhaps impossible to realize without the use of intensive cooling systems such as those discussed in \cite{dalir2022high_density}.

Each crossing, modulator, and filter within the $4\times4$ blocks has a certain amount of insertion loss. For a monolithic approach, as in the ``Scale-Up'' approach, the accumulation of these losses follows an exponential model, which eventually leads to a degradation of the optical Signal-to-Noise Ratio (SNR) below the threshold of the photodetector~\cite{al2022scaling}. As quantified in Section~\ref{sec:ptc4}, the combined insertion and fanout losses place the $4\times4$ chiplet at the practical boundary of the available link budget. The proposed approach, as in the ``Scale-Out'' approach, limits the size of the high-loss meshes to $4\times4$ while using an optical interposer for communication.

\subsection{Roadmap to Photonic Accelerator}
\label{sec:roadmap}

Optical signals propagate at the speed of light and do not possess a natural ``state-holding'' mechanism analogous to electronic SRAM or DRAM, implying that all persistent storage must currently reside in the electronic domain, with every data transfer to the photonic compute units incurring an E/O conversion penalty. An emerging approach to mitigate this limitation is the use of PCMs, which can encode weight values as refractive index states in a non-volatile manner~\cite{farmakidis2019plasmonic}, potentially reducing the volume of off-chip filter read operations that our results in Section~\ref{sec:memory} identify as a dominant bottleneck. However, PCM-based photonic memory remains at an early stage, with significant challenges in write endurance, multi-level programming precision, and integration density~\cite{shafiee2023compact}. A comprehensive memory architecture study must examine which data categories, including weights, activations, and partial sums, should reside in the photonic versus the electronic domain, and at what level of the hierarchy PCM or other photonic storage can be practically deployed.

Furthermore, should WDM technology mature sufficiently for deployment on photonic interposers, the throughput and parallelism of the scale-out architecture could be multiplied proportionally to the number of available wavelength channels,enabling the progression from 1$\lambda$ to multi-wavelength (4$\lambda$, 8$\lambda$) operation shown in Fig.~\ref{fig_pe_layout}. WDM-enabled broadcasting would allow simultaneous distribution of distinct data operands across multiple chiplets on separate wavelengths, effectively converting serial data transfers into parallel optical streams.

Recent work has addressed the challenge of scaling photonic circuits beyond current limitations from complementary directions: circuit-level strategies for managing analog signal degradation in extended photonic processing chains \cite{aadhi2025scalable}, mitigation of spurious wave interactions including crosstalk and back-reflections in large-scale systems \cite{shekhar2024roadmapping}, and high-density photonic interconnects for chip-to-chip communication \cite{tian2025photonic}. The architectural findings presented in this paper, particularly the ``Symmetric Grid Rule'' and the ``Utilization Wall'', provide system-level design guidelines that complement these device- and circuit-level advances. By combining the architectural guidelines established here with advances in device fabrication, the photonic research community can progress from proof-of-concept demonstrations toward production-grade accelerators competitive with today's electronic TPUs~\cite{jouppi2017datacenter}.

\section{Conclusion}
\label{sec:conclusion}

This paper addresses a critical gap in scalable photonic system-level architectures by proposing a modular Scale-Out paradigm based on $4\times4$ PTC blocks, explicitly accounting for the intrinsic physical limitations of photonic signal integrity and insertion loss. Through systematic architectural exploration across four representative DNN workloads and system sizes ranging from 64 to 1024 PEs, we establish two key findings. First, the \textit{Symmetric Grid Rule} demonstrates that symmetric grid topologies improve PE utilization by up to $6\times$ compared to linear configurations ($87.15\%$ versus $14.05\%$ for AlphaGoZero at 512 PEs) while reducing filter weight reads by up to $18\times$ under the same PE budget. Second, the \textit{Utilization Wall} reveals that topology dominates performance over raw hardware count: scaling GoogleNet from 128 to 1024 PEs yields only a $4.8\times$ speedup from an $8\times$ hardware increase ($\eta = 0.60$), with utilization declining from $85.29\%$ to $51.67\%$. Scaling robustness is workload-dependent, with AlphaGoZero maintaining $\eta = 0.78$ due to its uniform residual structure, while MobileNet degrades to $\eta = 0.42$. At the system level, the proposed photonic architecture achieves up to $11.3\times$ higher effective throughput than digital accelerators ($71.4f$ versus $6.3f$~ops/s) while maintaining competitive energy efficiency at 18.5 effective TOPS/W. These results provide architectural guidelines for the development of next-generation photonic chips and underscore the need for reconfigurable photonic interconnects that can adapt their grid topology to the heterogeneous demands of modern AI workloads.






\bibliographystyle{elsarticle-num} 
\bibliography{ref}

\end{document}

%% file: Figures_V03_Array/00_Matrix_V01.tex
\begin{tikzpicture}

\node[font=\small\bfseries, matA] at (0, 1.1) {Matrix $\mathbf{A}$ (Inputs)};
\matrix (matA) [matrix of math nodes,
    nodes={fill=matA!10, draw=white, line width=0.5pt, minimum width=0.8cm, minimum height=0.55cm, font=\small, anchor=center},
    row sep=-\pgflinewidth, column sep=-\pgflinewidth,
    draw=matA!60, line width=1pt, inner sep=0pt] at (0, 0)
{
    \textcolor{matA!80!black}{a_{11}} & \textcolor{matA!80!black}{a_{12}} & \textcolor{matA!80!black}{a_{13}} \\
    \textcolor{matA!80!black}{a_{21}} & \textcolor{matA!80!black}{a_{22}} & \textcolor{matA!80!black}{a_{23}} \\
    \textcolor{matA!80!black}{a_{31}} & \textcolor{matA!80!black}{a_{32}} & \textcolor{matA!80!black}{a_{33}} \\
};

\node[font=\Large] at (1.8, 0) {$\times$};

\node[font=\small\bfseries, matB] at (3.6, 1.1) {Matrix $\mathbf{B}$ (Weights)};
\matrix (matB) [matrix of math nodes,
    nodes={fill=matB!10, draw=white, line width=0.5pt, minimum width=0.8cm, minimum height=0.55cm, font=\small, anchor=center},
    row sep=-\pgflinewidth, column sep=-\pgflinewidth,
    draw=matB!60, line width=1pt, inner sep=0pt] at (3.6, 0)
{
    \textcolor{matB!80!black}{b_{11}} & \textcolor{matB!80!black}{b_{12}} & \textcolor{matB!80!black}{b_{13}} \\
    \textcolor{matB!80!black}{b_{21}} & \textcolor{matB!80!black}{b_{22}} & \textcolor{matB!80!black}{b_{23}} \\
    \textcolor{matB!80!black}{b_{31}} & \textcolor{matB!80!black}{b_{32}} & \textcolor{matB!80!black}{b_{33}} \\
};

\node[font=\Large] at (5.4, 0) {$=$};

\node[font=\small\bfseries, matC] at (7.2, 1.1) {Matrix $\mathbf{C}$ (Output)};
\matrix (matC) [matrix of math nodes,
    nodes={fill=matC!10, draw=white, line width=0.5pt, minimum width=0.8cm, minimum height=0.55cm, font=\small, anchor=center},
    row sep=-\pgflinewidth, column sep=-\pgflinewidth,
    draw=matC!60, line width=1pt, inner sep=0pt] at (7.2, 0)
{
    \textcolor{matC!80!black}{c_{11}} & \textcolor{matC!80!black}{c_{12}} & \textcolor{matC!80!black}{c_{13}} \\
    \textcolor{matC!80!black}{c_{21}} & \textcolor{matC!80!black}{c_{22}} & \textcolor{matC!80!black}{c_{23}} \\
    \textcolor{matC!80!black}{c_{31}} & \textcolor{matC!80!black}{c_{32}} & \textcolor{matC!80!black}{c_{33}} \\
};

\node[font=\footnotesize, text=wireGray, align=center] at (3.6, -1.2) {%
$c_{ij} = \sum_{k=1}^{3} a_{ik}\cdot b_{kj}$ \quad $\Rightarrow$ \quad \textbf{27 multiplications} + 18 additions};

\end{tikzpicture}

%% file: Figures_V03_Array/01_Electronic_V01.tex
\begin{tikzpicture}[
    >=Stealth,
    every node/.style={}, 
    pe/.style={draw=systGray, fill=systGray!15, minimum size=0.9cm, rounded corners=2pt, font=\small, line width=0.6pt}
]


\foreach \i in {0,1,2} {
    \foreach \j in {0,1,2} {
        \node[pe] (pe\i\j) at (1.8+\j*1.6, 8.5-\i*1.6) {PE$_{\pgfmathparse{int(\i+1)}\pgfmathresult\pgfmathparse{int(\j+1)}\pgfmathresult}$};
    }
}

\foreach \i in {0,1,2} {
    \draw[->, matA, line width=0.7pt] (pe\i0.west)+(-0.55,0) -- (pe\i0.west);
    \node[font=\footnotesize, matA, left] at ($(pe\i0.west)+(-0.55,0)$) {$a_{\pgfmathparse{int(\i+1)}\pgfmathresult,k}$};
    \foreach \j [evaluate=\j as \jn using int(\j+1)] in {0,1} {
        \draw[->, matA!60, line width=0.5pt] (pe\i\j.east) -- (pe\i\jn.west);
    }
    \draw[->, matA!40, line width=0.3pt] (pe\i2.east) -- ++(0.35,0);
}

\foreach \j in {0,1,2} {
    \draw[->, matB, line width=0.7pt] (pe0\j.north)+(0,0.55) -- (pe0\j.north);
    \node[font=\footnotesize, matB, above] at ($(pe0\j.north)+(0,0.55)$) {$b_{k,\pgfmathparse{int(\j+1)}\pgfmathresult}$};
    \foreach \i [evaluate=\i as \iin using int(\i+1)] in {0,1} {
        \draw[->, matB!60, line width=0.5pt] (pe\i\j.south) -- (pe\iin\j.north);
    }
    \draw[->, matB!40, line width=0.3pt] (pe2\j.south) -- ++(0,-0.35);
}

\foreach \j [evaluate=\j as \jj using int(\j+1)] in {0,1,2} {
    \node[font=\footnotesize, matC] at (1.8+\j*1.6, 4.35) {$c_{i,\jj}$};
}


\node[draw=systGray!70, fill=systGray!5, rounded corners=4pt, dashed, line width=0.6pt, minimum width=4cm, minimum height=2.2cm] (pebox) at (8.0, 6.9) {};

\draw[->, systGray!80, dashed, line width=0.8pt] (pe12.north east) ++(0.05, 0.05) to[out=30, in=180] (pebox.west);

\begin{scope}[shift={(6.85, 6.4)}]
    \draw[->, matA!50, line width=0.6pt] (-0.1, 0.8) node[left, font=\footnotesize, matA!80!black] {$a_{in}$} -- (2.6, 0.8) node[right, font=\footnotesize, matA!80!black] {$a_{out}$};
    \draw[->, matB!50, line width=0.6pt] (0.8, 1.2) node[above, font=\footnotesize, matB!80!black] {$b_{in}$} -- (0.8, -0.2) node[below, font=\footnotesize, matB!80!black] {$b_{out}$};
    
    \draw[->, matA, line width=0.5pt] (0.5, 0.8) |- (0.65, 0.4);
    \fill[matA!50] (0.5, 0.8) circle (0.8pt);
    
    \draw[->, matB, line width=0.5pt] (0.8, 0.8) -- (0.8, 0.55);
    \fill[matB!50] (0.8, 0.8) circle (0.8pt);
    
    \node[draw=systGray, fill=white, circle, inner sep=0pt, minimum size=0.35cm, font=\footnotesize] (mult) at (0.8, 0.4) {$\times$};
    \node[draw=systGray, fill=white, circle, inner sep=0pt, minimum size=0.35cm, font=\footnotesize] (add) at (1.45, 0.4) {$+$};
    \node[draw=systGray, fill=systGray!15, minimum width=0.45cm, minimum height=0.35cm, font=\footnotesize, rounded corners=1pt] (acc) at (2.15, 0.4) {Acc};
    
    \draw[->, systGray, line width=0.5pt] (mult.east) -- (add.west);
    \draw[->, systGray, line width=0.5pt] (add.east) -- (acc.west);
    \draw[->, systGray, line width=0.5pt] (acc.south) -- ++(0,-0.15) -| (add.south);
\end{scope}


\begin{scope}[shift={(1.3, 2.5)}]
    \draw[systGray!70, line width=0.5pt, decorate, decoration={snake, amplitude=1.2pt, segment length=5pt}] (0.3, 1.1) -- (7.2, 1.1);
    \node[font=\footnotesize, systGray] at (3.75, 0.75) {Clock: data shifts one PE per cycle};

    \foreach \t [count=\ti from 0] in {1,2,3,4,5} {
        \draw[systGray!50, line width=0.3pt] (0.6+\ti*1.1, 0.4) -- (0.6+\ti*1.1, -0.3);
        \node[font=\footnotesize, systGray!70] at (0.6+\ti*1.1, -0.6) {$t_{\t}$};
    }
    \node[font=\footnotesize, matA, fill=matA!10, rounded corners=1pt, inner sep=1.5pt] at (1.15, -0.2) {$a_{11}$};
    \node[font=\footnotesize, matA, fill=matA!10, rounded corners=1pt, inner sep=1.5pt] at (2.25, -0.2) {$a_{12}$};
    \node[font=\footnotesize, matA, fill=matA!10, rounded corners=1pt, inner sep=1.5pt] at (3.35, -0.2) {$a_{13}$};
    \draw[->, systGray!50, line width=0.4pt] (1.5, -0.2) -- (1.9, -0.2);
    \draw[->, systGray!50, line width=0.4pt] (2.6, -0.2) -- (3.0, -0.2);
    \node[font=\footnotesize, systGray!80, align=center] at (5.8, -0.2) {$2n\!-\!1 = 5$\\[-1pt]cycles};
\end{scope}

\end{tikzpicture}

%% file: Figures_V03_Array/02_Photonic_V01.tex
\begin{tikzpicture}[
    >=Stealth,
    every node/.style={font=\rmfamily}, 
    mzi/.style={draw=mziCol!90!black, fill=mziCol!10, minimum width=0.52cm, minimum height=0.32cm, rounded corners=1.5pt, font=\tiny, line width=0.5pt},
    dacbox/.style={draw=inCol!90!black, fill=inCol!10, minimum width=0.45cm, minimum height=0.3cm, rounded corners=1.5pt, font=\tiny\bfseries, line width=0.5pt},
    drvbox/.style={draw=inCol!90!black, fill=inCol!20, minimum width=0.45cm, minimum height=0.3cm, rounded corners=1.5pt, font=\tiny\bfseries, line width=0.5pt},
    adcbox/.style={draw=outCol!90!black, fill=outCol!20, minimum width=0.5cm, minimum height=0.3cm, rounded corners=1.5pt, font=\tiny\bfseries, line width=0.5pt},
    pdbox/.style={draw=outCol!90!black, fill=outCol!10, minimum width=0.38cm, minimum height=0.3cm, rounded corners=1.5pt, font=\tiny\bfseries, line width=0.5pt},
    tiabox/.style={draw=outCol!90!black, fill=outCol!15, minimum width=0.42cm, minimum height=0.3cm, rounded corners=1.5pt, font=\tiny\bfseries, line width=0.5pt},
    laserbox/.style={draw=matB!90!black, fill=matB!10, minimum width=0.5cm, minimum height=0.3cm, rounded corners=1.5pt, font=\tiny\bfseries, line width=0.5pt},
    modbox/.style={draw=matB!90!black, fill=matB!25, minimum width=0.5cm, minimum height=0.3cm, rounded corners=1.5pt, font=\tiny\bfseries, line width=0.5pt},
    inbox/.style={draw=matA!90!black, fill=matA!10, minimum width=0.65cm, minimum height=0.35cm, rounded corners=1.5pt, font=\tiny\bfseries, line width=0.5pt},
    outbox/.style={draw=matC!90!black, fill=matC!10, minimum width=0.65cm, minimum height=0.35cm, rounded corners=1.5pt, font=\tiny\bfseries, line width=0.5pt},
    wg/.style={wgColor, line width=0.9pt},
    vbus/.style={matC, line width=1.1pt},
    combdot/.style={fill=matC, circle, inner sep=1.2pt}
]

\foreach \j [evaluate=\j as \jlabel using int(\j+1),
             evaluate=\j as \xj using {11.8+\j*2.5}] in {0,1,2} {
    
    \node[font=\tiny\bfseries, matB!90!black] at (\xj-0.9, 10.28) {col $j\!=\!\jlabel$};

    \foreach \k [evaluate=\k as \klabel using int(\k+1),
                 evaluate=\k as \xin using {\xj - 1.4 + \k*0.45},
                 evaluate=\k as \yend using {0.85 - \k*0.8}] in {0,1,2} {
        
        \node[font=\tiny, matB!90!black] (b\k\j) at (\xin, 10) {$b_{\klabel\jlabel}$};
        \node[inbox, draw=matB!90!black, fill=matB!10, rotate=-90] (bin\j\k) at (\xin, 9.2) {Input};
        \draw[->, wireGray, line width=0.4pt] (b\k\j.south) -- (bin\j\k.west);
        
        \draw[wg, matB] (bin\j\k.east) -- (\xin, \yend);
    }
}

\foreach \row [evaluate=\row as \rowlabel using int(\row+1),
               evaluate=\row as \yc using {6.9-\row*3.6}] in {0,1,2} {

    \node[font=\tiny\bfseries, matA!90!black, rotate=90] at (8.3,\yc) {Row \rowlabel};

    \foreach \k [evaluate=\k as \klabel using int(\k+1),
                 evaluate=\k as \yk using {\yc+0.8-\k*0.8}] in {0,1,2} {
        
        \node[font=\tiny, matA!90!black] (ain) at (8.7, \yk+0.28) {$a_{\rowlabel\klabel}$};
        \node[inbox] (in\row\k) at (9.6, \yk+0.28) {Input};
        \draw[->, wireGray, line width=0.4pt] (ain.east) -- (in\row\k.west);
        \draw[wg, matA] (in\row\k.east) -- (16.4, \yk+0.28);
        
        \foreach \j [evaluate=\j as \jlabel using int(\j+1),
                     evaluate=\j as \xj using {11.8+\j*2.5}] in {0,1,2} {
            
            \pgfmathsetmacro{\bx}{\xj - 1.4 + \k*0.45}

            \node[mzi] (mzi\row\k\j) at (\xj,\yk-0.1) {};
            
            \draw[wg, matA, rounded corners=3pt] (\xj-0.4, \yk+0.28) -| (mzi\row\k\j.north);
            \fill[matA] (\xj-0.4, \yk+0.28) circle (0.8pt);

            \draw[wg, matB, rounded corners=3pt] (\bx, \yk+0.35) -- (\bx, \yk-0.1) -- (mzi\row\k\j.west);
            \fill[matB] (\bx, \yk+0.35) circle (0.8pt);
        }
    }
    
    \foreach \j [evaluate=\j as \jlabel using int(\j+1),
                 evaluate=\j as \xj using {11.8+\j*2.5}] in {0,1,2} {
        
        \draw[wg, vbus] (\xj+0.5, \yc+0.6) -- (\xj+0.5, \yc-1.1);
        
        \foreach \k [evaluate=\k as \yk using {\yc+0.8-\k*0.8}] in {0,1,2} {
            \draw[wg, matC, rounded corners=3pt] (mzi\row\k\j.east) -| (\xj+0.5, \yk-0.2);
            \node[combdot] at (\xj+0.5, \yk-0.2) {};
        }
        
        \node[font=\tiny, matC!90!black, fill=white, inner sep=1pt] at (\xj+0.5, \yc-0.95) {$\Sigma$};
        
        \node[outbox] (out\row\j) at (\xj+0.5, \yc-1.45) {Output};
        \draw[->, wireGray, line width=0.4pt] (\xj+0.5, \yc-1.1) -- (out\row\j.north);
        
        \draw[->, wireGray, line width=0.4pt] (out\row\j.south) -- ++(0,-0.25) node[below, font=\tiny, matC!90!black, inner sep=1pt] {$c_{\rowlabel\jlabel}$};
    }
    
    \pgfmathparse{int(\row)}
    \ifnum\pgfmathresult<2
        \draw[wireGray!50, dashed, line width=0.4pt] (8.5,\yc-2.25) -- (17.8,\yc-2.25);
    \fi
}

\begin{scope}[yshift=-0.75cm] 
    \draw[wireGray!60, rounded corners=3pt, line width=0.4pt] (8.4,-1.8) rectangle (15.8,-3.2);
    \node[font=\tiny\bfseries, black!85] at (9.1,-2.1) {Signal Chain:};

    \node[dacbox, minimum width=0.3cm, minimum height=0.22cm] (sc_dac) at (9.0,-2.7) {\tiny DAC};
    \node[font=\tiny] at (9.35,-2.7) {$\rightarrow$};
    \node[drvbox, minimum width=0.3cm, minimum height=0.22cm] (sc_drv) at (9.7,-2.7) {\tiny Drv};
    \node[font=\tiny] at (10.05,-2.7) {$\rightarrow$};
    \node[laserbox, minimum width=0.3cm, minimum height=0.22cm] (sc_las) at (10.45,-2.7) {\tiny Las};
    \node[font=\tiny] at (10.8,-2.7) {$\rightarrow$};
    \node[modbox, minimum width=0.3cm, minimum height=0.22cm] (sc_mod) at (11.2,-2.7) {\tiny Mod};

    \node[draw=inCol!90!black, dashed, rounded corners=2pt, fit=(sc_dac) (sc_mod), inner sep=2pt] (sc_in_fit) {};
    \node[font=\tiny\bfseries, inCol!90!black, above=1pt of sc_in_fit, xshift=15pt] {"Input" Module};

    \node[font=\tiny] at (11.6,-2.7) {$\rightarrow$};
    \node[font=\tiny, matA!90!black] at (11.85,-2.7) {1:3};
    \node[font=\tiny] at (12.1,-2.7) {$\rightarrow$};
    \node[mzi, minimum width=0.3cm, minimum height=0.22cm] at (12.45,-2.7) {};
    \node[font=\tiny] at (12.8,-2.7) {$\rightarrow$};
    \node[font=\tiny, matC!90!black] at (13.0,-2.7) {$\Sigma$};
    \node[font=\tiny] at (13.2,-2.7) {$\rightarrow$};

    \node[pdbox, minimum width=0.28cm, minimum height=0.22cm] (sc_pd) at (13.55,-2.7) {\tiny PD};
    \node[font=\tiny] at (13.85,-2.7) {$\rightarrow$};
    \node[tiabox, minimum width=0.3cm, minimum height=0.22cm] (sc_tia) at (14.2,-2.7) {\tiny TIA};
    \node[font=\tiny] at (14.55,-2.7) {$\rightarrow$};
    \node[adcbox, minimum width=0.3cm, minimum height=0.22cm] (sc_adc) at (14.95,-2.7) {\tiny ADC};

    \node[draw=outCol!90!black, dashed, rounded corners=2pt, fit=(sc_pd) (sc_adc), inner sep=2pt] (sc_out_fit) {};
    \node[font=\tiny\bfseries, outCol!90!black, above=-12pt of sc_out_fit, xshift=45pt] {"Output" Module};

    \draw[->, inCol!90!black, dashed, line width=0.6pt] (in22.south) to[out=270, in=90] (sc_in_fit.north);
    \draw[->, outCol!90!black, dashed, line width=0.6pt] (out22.south) to[out=270, in=90] (sc_out_fit.north);
    

\end{scope}

\end{tikzpicture}

%% file: Figures/03_ScaleUp_Npe.tex
\begin{tikzpicture}[
    >=Stealth,
    node distance=0.5cm,
    every node/.style={font=\rmfamily},
    digblock/.style={
        rectangle, rounded corners=3pt,
        draw=blockborder, fill=blockfill,
        text=eictext, font=\rmfamily\small,
        minimum width=4.4cm, minimum height=1.0cm,
        align=center, line width=0.6pt
    },
    phoblock/.style={
        rectangle, rounded corners=3pt,
        draw=blockborder, fill=blockfill,
        text=pictext, font=\rmfamily\small,
        minimum width=6.5cm,
        minimum height=0.9cm,
        align=center, line width=0.6pt
    },
    peblock/.style={
        rectangle, rounded corners=2pt,
        draw=peborder, fill=pefill,
        text=petext, font=\rmfamily\footnotesize\bfseries,
        minimum width=1.0cm, minimum height=1.0cm,
        align=center, line width=0.6pt
    },
    arrdig/.style={-Stealth, line width=0.7pt, color=blockborder, rounded corners=3pt},
    arrpho/.style={-Stealth, line width=0.7pt, color=blockborder, rounded corners=3pt},
    arract/.style={-Stealth, line width=1.0pt, color=actarrow, rounded corners=4pt}, 
    arrwt/.style={-Stealth, line width=1.0pt, color=wtarrow, rounded corners=4pt, dash pattern=on 4pt off 2pt}, 
    arrret/.style={-Stealth, line width=1.0pt, color=photoarrow, rounded corners=4pt, dash pattern=on 4pt off 2pt on 1pt off 2pt}, 
    chiplabel/.style={font=\rmfamily\bfseries\normalsize, color=black},
    dotsstyle/.style={font=\Large\bfseries, color=black!50},
]


\node[digblock] (mem) 
    {\textbf{Input Memory (SRAM)}\\[1pt]{\scriptsize Weights + Input}};
\node[digblock, below=0.4cm of mem] (dac) 
    {\textbf{DAC}\\[1pt]{\scriptsize Digital $\rightarrow$ Analog}};

\node[digblock, below=5.2cm of dac] (tia) 
    {\textbf{TIA}\\[1pt]{\scriptsize Current $\rightarrow$ Voltage}};
\node[digblock, below=0.4cm of tia] (adc) 
    {\textbf{ADC}\\[1pt]{\scriptsize Analog $\rightarrow$ Digital}};

\node[digblock, below=0.4cm of adc] (outmem) 
    {\textbf{Output Memory (SRAM)}};
\node[digblock, below=0.4cm of outmem] (digfunc) 
    {\textbf{Digital Functions}\\[1pt]{\scriptsize ReLU, Pooling, Normalization}};

\draw[arrdig] (mem) -- (dac);
\draw[arrdig] (tia) -- (adc);
\draw[arrdig] (adc) -- (outmem);
\draw[arrdig] (outmem) -- (digfunc);
\draw[arrdig, dash pattern=on 3pt off 2pt] (digfunc.west) -- ++(-0.6,0) |- (mem.west);
\node[font=\rmfamily\Large, color=blockborder, rotate=90, anchor=south] 
    at ($(digfunc.west)+(-0.85, 2.5)$) {Next layer};


\node[phoblock, right=4.5cm of mem, minimum width=4.5cm, minimum height=0.75cm] (laser) 
    {\textbf{Laser Source (CW)}};
\node[phoblock, below=0.35cm of laser, minimum width=4.5cm] (mod) 
    {\textbf{Optical Modulator}\\[1pt]{\scriptsize E/O: MZI\,/\,EAM\,/\,SLM}};
\node[phoblock, below=0.35cm of mod, minimum height=0.75cm, minimum width=6.5cm] (wg) 
    {\textbf{Routing}\\[1pt]{\scriptsize Waveguide / Splitters / MRR}};

\node[peblock, below=0.6cm of wg, xshift=-2.8cm] (pe0) {PE$_0$};
\node[peblock, right=0.2cm of pe0, yshift=-0.95cm] (pe1) {PE$_1$};
\node[dotsstyle, right=0.45cm of pe1, yshift=-0.95cm] (dots) {$\ddots$};
\node[peblock, right=0.45cm of dots, yshift=-2.5cm] (pen1) {PE$_{n\text{-}1}$};
\node[peblock, right=0.2cm of pen1, yshift=-0.95cm] (pen) {PE$_n$};

\coordinate (combiner_y) at ($(pen.south)+(0, -0.5)$);
\node[phoblock, minimum height=0.75cm, minimum width=6.5cm] (combiner) at (wg |- combiner_y) 
    {\textbf{Output Combiner}};
\node[phoblock, below=0.35cm of combiner, minimum width=4.5cm] (pd) 
    {\textbf{Photodetector (PD)}\\[1pt]{\scriptsize O/E: Optical $\rightarrow$ Current}};

\draw[arrpho] (laser) -- (mod);
\draw[arrpho] (mod) -- (wg);
\foreach \pe in {pe0, pe1, pen1, pen}{
    \draw[arrpho] (wg.south -| \pe.north) -- (\pe.north);
}
\draw[arrpho, dash pattern=on 2pt off 2pt] (wg.south -| dots.north) -- (dots.north);
\foreach \pe in {pe0, pe1, pen1, pen}{
    \draw[arrpho] (\pe.south) -- (\pe.south |- combiner.north);
}
\draw[arrpho, dash pattern=on 2pt off 2pt] (dots.south) -- (dots.south |- combiner.north);
\draw[arrpho] (combiner) -- (pd);


\coordinate (IF) at ($(mem.east)!0.5!(laser.west)$);

\draw[arract] 
    ($(dac.south)+(0.15, 0)$) -- ($(dac.south)+(0.15, -2.3)$)
    -| ($(dac.south -| IF)+(0.3, 0)$)
    |- (mod.west);

\coordinate (bracket_top) at ($(pe0.west)+(-0.5, 0)$);
\coordinate (bracket_bot) at (bracket_top |- pen.west);
\coordinate (bracket_mid) at ($(bracket_top)!0.5!(bracket_bot)$);

\draw[arrwt] 
    ($(dac.south)+(-0.15, 0)$) -- ($(dac.south)+(-0.15, -2.6)$)
    -| ($(dac.south -| IF)+(0.35, -2.9)$)
    |- (bracket_mid);

\draw[wtarrow, line width=1.0pt] 
    ($(bracket_top)+(0.12, 0)$) -- (bracket_top) 
    -- (bracket_bot) -- ($(bracket_bot)+(0.12, 0)$);

\foreach \pe in {pe0, pe1, pen1, pen}{
    \draw[arrwt, line width=0.7pt] (bracket_top |- \pe.west) -- (\pe.west);
}
\draw[wtarrow, line width=0.7pt, dash pattern=on 2pt off 2pt] 
    (bracket_top |- dots.west) -- (dots.west);

\draw[arrret] 
    (pd.west) -- ($(pd.west -| IF)+(0.2, 0)$)
    |- ($(tia.north)+(0, 2.3)$) -- (tia.north);


\begin{scope}[on background layer]
    \coordinate (P1) at ($(dac.west)+(-0.35, 0)$);
    \coordinate (P2) at (P1 -| IF);
    \coordinate (P3) at ($(IF |- laser.north) + (0, 0.35)$);
    \coordinate (P4) at ($(wg.east |- P3) + (0.35, 0)$);
    \coordinate (P5) at ($(P4 |- pd.south) + (0, -0.35)$);
    \coordinate (P6) at (P5 -| IF);
    \coordinate (P7) at (adc.west -| IF);
    \coordinate (P8) at (P1 |- P7);

    \fill[analogfill, fill opacity=0.30, rounded corners=8pt]
        (P1) -- (P2) -- (P3) -- (P4) -- (P5) -- (P6) -- (P7) -- (P8) -- cycle;
    \draw[analogborder, line width=1.2pt, dash pattern=on 6pt off 3pt, rounded corners=8pt]
        (P1) -- (P2) -- (P3) -- (P4) -- (P5) -- (P6) -- (P7) -- (P8) -- cycle;
\end{scope}


\begin{scope}[on background layer]
    \coordinate (DT1) at ($(mem.north west)+(-0.2, 0.2)$);
    \coordinate (DT2) at ($(dac.east)+(0.2, 0)$);
    \fill[digitalfill, fill opacity=0.40, rounded corners=5pt]
        (DT1) rectangle (DT2);
    \draw[digitalborder, line width=1.0pt, dash pattern=on 2pt off 2pt, rounded corners=5pt]
        (DT1) rectangle (DT2);
\end{scope}

\begin{scope}[on background layer]
    \coordinate (DB1) at ($(adc.west)+(-0.2, 0)$);
    \coordinate (DB2) at ($(digfunc.south east)+(0.2, -0.2)$);
    \fill[digitalfill, fill opacity=0.40, rounded corners=5pt]
        (DB1) rectangle (DB2);
    \draw[digitalborder, line width=1.0pt, dash pattern=on 2pt off 2pt, rounded corners=5pt]
        (DB1) rectangle (DB2);
\end{scope}


\begin{scope}[on background layer]
    \node[rectangle, rounded corners=8pt, 
          draw=eicborder, line width=1.4pt, 
          fill=eicfill, fill opacity=0.3,
          inner sep=14pt,
          fit=(mem)(dac)(tia)(adc)(outmem)(digfunc), 
          label={[chiplabel, color=eicborder]above:{Electronic Chip (EIC)}}] (eic) {};
          
    \coordinate (pic_left)  at ([xshift=-39pt]pe0.west);   
    \coordinate (pic_right) at ([xshift=14pt]wg.east);     
    \coordinate (pic_top)   at ([yshift=15pt]laser.north); 
    \coordinate (pic_bot)   at ([yshift=-20pt]pd.south);   

    \node[rectangle, rounded corners=8pt, 
          draw=picborder, line width=1.4pt, 
          fill=picfill, fill opacity=0.3,
          inner sep=0pt, 
          fit=(pic_left)(pic_right)(pic_top)(pic_bot), 
          label={[chiplabel, color=picborder]above:{Photonic Chip (PIC)}}] (pic) {};
\end{scope}

\node[font=\rmfamily\scriptsize, color=chipdash, rotate=90, anchor=south] 
    at ($(eic.east)!0.5!(pic.west)+(0.2, -0.8)$) {Chip-to-chip interface};


\coordinate (legend_anchor) at ($(IF |- digfunc.south) + (0, -1.0)$);

\node[font=\rmfamily\small, anchor=west, color=black] 
    (leg_ana_text) at ($(legend_anchor) + (-5.0, 0)$) {Analog Domain};
\node[rectangle, rounded corners=2pt, draw=analogborder, line width=1.0pt, 
      dash pattern=on 6pt off 3pt, fill=analogfill, fill opacity=0.45,
      minimum width=0.55cm, minimum height=0.35cm,
      anchor=east] at ($(leg_ana_text.west) + (-0.15, 0)$) {};

\node[font=\rmfamily\small, anchor=west, color=black] 
    (leg_dig_text) at ($(leg_ana_text.south west) + (0, -0.6)$) {Digital Domain};
\node[rectangle, rounded corners=2pt, draw=digitalborder, line width=1.0pt, 
      dash pattern=on 2pt off 2pt, fill=digitalfill, fill opacity=0.40,
      minimum width=0.55cm, minimum height=0.35cm,
      anchor=east] at ($(leg_dig_text.west) + (-0.15, 0)$) {};

\node[font=\rmfamily\small, anchor=west, color=black] 
    (leg_eic_text) at ($(legend_anchor) + (-0.5, 0)$) {Electronic Chip (EIC)};
\node[rectangle, rounded corners=2pt, draw=eicborder, line width=1.0pt, 
      fill=eicfill, fill opacity=0.5, minimum width=0.55cm, minimum height=0.35cm,
      anchor=east] at ($(leg_eic_text.west) + (-0.15, 0)$) {};

\node[font=\rmfamily\small, anchor=west, color=black] 
    (leg_pic_text) at ($(leg_eic_text.south west) + (0, -0.6)$) {Photonic Chip (PIC)};
\node[rectangle, rounded corners=2pt, draw=picborder, line width=1.0pt, 
      fill=picfill, fill opacity=0.5, minimum width=0.55cm, minimum height=0.35cm,
      anchor=east] at ($(leg_pic_text.west) + (-0.15, 0)$) {};

\node[font=\rmfamily\small, anchor=west, color=black] 
    (leg_act_text) at ($(legend_anchor) + (4.0, 0)$) {Input};
\draw[arract] ($(leg_act_text.west) + (-0.8, 0)$) -- ($(leg_act_text.west) + (-0.15, 0)$);

\node[font=\rmfamily\small, anchor=west, color=black] 
    (leg_wt_text) at ($(leg_act_text.south west) + (0, -0.4)$) {Weights};
\draw[arrwt] ($(leg_wt_text.west) + (-0.8, 0)$) -- ($(leg_wt_text.west) + (-0.15, 0)$);

\node[font=\rmfamily\small, anchor=west, color=black] 
    (leg_pho_text) at ($(leg_wt_text.south west) + (0, -0.4)$) {$I_{\mathrm{photo}}$};
\draw[arrret] ($(leg_pho_text.west) + (-0.8, 0)$) -- ($(leg_pho_text.west) + (-0.15, 0)$);

\end{tikzpicture}

%% file: Figures_V02_Interposers/01_Electronic_V01.tex
\begin{tikzpicture}[
    >=Stealth,
    chipbox/.style={
        rectangle, rounded corners=6pt, 
        draw=chipborder, fill=chipfill, 
        text=chiptxt, font=\sffamily\small\bfseries, 
        minimum width=2.4cm, minimum height=1.6cm, 
        align=center, line width=1.2pt
    },
    routernode/.style={
        rectangle, fill=routerfill, draw=routerborder, 
        line width=0.8pt, rounded corners=2pt, 
        inner sep=2pt, font=\sffamily\tiny\bfseries
    },
    meshline/.style={
        line width=2.5pt, color=meshbus
    },
    meshconn/.style={
        -Stealth, line width=1.2pt, color=buscolor
    },
    dotlabel/.style={
        font=\Huge\bfseries, color=buscolor
    },
    drambox/.style={
        rectangle, rounded corners=6pt,
        draw=dramborder, fill=dramfill,
        text=chiptxt, font=\sffamily\large\bfseries,
        minimum width=8.5cm, minimum height=1.2cm,
        align=center, line width=1.5pt
    }
]

\def\dx{4.0}   
\def\dy{-3.2}  
\def\busY{1.2} 

\node[drambox] (DRAM) at (1.5*\dx, 2.2) {Off-Chip Memory (DRAM)};

\node[chipbox] (C00) at (0, 0)       {Chiplet\\$(0,0)$};
\node[chipbox] (C01) at (\dx, 0)     {Chiplet\\$(0,1)$};
\node[dotlabel] at (2*\dx, 0) {$\cdots$};
\node[chipbox] (C0n) at (3*\dx, 0)   {Chiplet\\$(0,n)$};

\draw[meshconn, <->] (DRAM.south) -- (1.5*\dx, \busY) -| (C00.north);
\draw[meshconn, <->] (DRAM.south) -- (1.5*\dx, \busY) -| (C01.north);
\draw[meshconn, <->] (DRAM.south) -- (1.5*\dx, \busY) -| (C0n.north);

\node[chipbox] (C10) at (0, \dy)       {Chiplet\\$(1,0)$};
\node[chipbox] (C11) at (\dx, \dy)     {Chiplet\\$(1,1)$};
\node[dotlabel] at (2*\dx, \dy) {$\cdots$};
\node[chipbox] (C1n) at (3*\dx, \dy)   {Chiplet\\$(1,n)$};

\node[dotlabel] at (0, 2*\dy)        {$\vdots$};
\node[dotlabel] at (\dx, 2*\dy)      {$\vdots$};
\node[dotlabel] at (2*\dx, 2*\dy)    {$\ddots$};
\node[dotlabel] at (3*\dx, 2*\dy)    {$\vdots$};

\node[chipbox] (Cm0) at (0, 3*\dy)       {Chiplet\\$(m,0)$};
\node[chipbox] (Cm1) at (\dx, 3*\dy)     {Chiplet\\$(m,1)$};
\node[dotlabel] at (2*\dx, 3*\dy) {$\cdots$};
\node[chipbox] (Cmn) at (3*\dx, 3*\dy)   {Chiplet\\$(m,n)$};

\foreach \node/\pos in {C00/above left, C01/above left, C0n/above right, C10/above left, C11/above left, C1n/above right, Cm0/below left, Cm1/below left, Cmn/below right} {
    \node[routernode, \pos=0.05cm of \node] (R\node) {R};
}

\draw[meshconn, <->] (C00.east) -- (C01.west);
\draw[meshconn, ->] (C01.east) -- ++(0.6,0);
\draw[meshconn, <-] (C0n.west) -- ++(-0.6,0);

\draw[meshconn, <->] (C10.east) -- (C11.west);
\draw[meshconn, ->] (C11.east) -- ++(0.6,0);
\draw[meshconn, <-] (C1n.west) -- ++(-0.6,0);

\draw[meshconn, <->] (Cm0.east) -- (Cm1.west);
\draw[meshconn, ->] (Cm1.east) -- ++(0.6,0);
\draw[meshconn, <-] (Cmn.west) -- ++(-0.6,0);

\draw[meshconn, <->] (C00.south) -- (C10.north);
\draw[meshconn, ->] (C10.south) -- ++(0,-0.5);
\draw[meshconn, <-] (Cm0.north) -- ++(0,0.5);

\draw[meshconn, <->] (C01.south) -- (C11.north);
\draw[meshconn, ->] (C11.south) -- ++(0,-0.5);
\draw[meshconn, <-] (Cm1.north) -- ++(0,0.5);

\draw[meshconn, <->] (C0n.south) -- (C1n.north);
\draw[meshconn, ->] (C1n.south) -- ++(0,-0.5);
\draw[meshconn, <-] (Cmn.north) -- ++(0,0.5);

\begin{scope}[on background layer]
    \foreach \r in {0, \dy, 3*\dy} {
        \draw[meshline] (-1.8, \r) -- (3*\dx+1.8, \r);
    }
    \foreach \c in {0, \dx, 3*\dx} {
        \draw[meshline] (\c, 1.2) -- (\c, 3*\dy-1.2);
    }
\end{scope}

\end{tikzpicture}

%% file: Figures_V02_Interposers/02_Photonics_V01.tex
\begin{tikzpicture}[>=Stealth, font=\rmfamily]

\def\wgL{0}      
\def\wgR{12}     
\def\urad{1.5}   

\def\rA{3.0}     
\def\rB{0.0}     
\def\rC{-9.0}    

\draw[line width=8pt, color=wgclad, line cap=round]
    (\wgL, \rA) -- (\wgR, \rA)
    arc[start angle=90, end angle=-90, radius=\urad]
    -- (\wgL, \rB);

\draw[line width=8pt, color=wgclad!60, line cap=round, dashed]
    (\wgL, \rB)
    arc[start angle=90, end angle=180, radius=\urad]
    -- (\wgL-\urad, \rC+\urad)
    arc[start angle=180, end angle=270, radius=\urad]
    -- (\wgL+2.5, \rC);

\draw[line width=8pt, color=wgclad, line cap=round]
    (\wgL+2.5, \rC) -- (\wgR, \rC);

\draw[line width=8pt, color=wgclad, line cap=round] 
    (\wgR, \rC) -- (\wgR+2.5, \rC);
\draw[line width=8pt, color=wgclad, line cap=round] 
    (-2.5, \rA) -- (\wgL, \rA);

\draw[line width=2.5pt, color=wgcore, line cap=round]
    (\wgL, \rA) -- (\wgR, \rA)
    arc[start angle=90, end angle=-90, radius=\urad]
    -- (\wgL, \rB);

\draw[line width=2.5pt, color=wgcore, dashed, opacity=0.6]
    (\wgL, \rB)
    arc[start angle=90, end angle=180, radius=\urad]
    -- (\wgL-\urad, \rC+\urad)
    arc[start angle=180, end angle=270, radius=\urad]
    -- (\wgL+2.5, \rC);

\draw[line width=2.5pt, color=wgcore, line cap=round]
    (\wgL+2.5, \rC) -- (\wgR, \rC);

\draw[line width=2.5pt, color=wgcore, line cap=round] 
    (-2.5, \rA) -- (\wgL, \rA);
\draw[line width=2.5pt, color=wgcore, line cap=round] 
    (\wgR, \rC) -- (\wgR+2.5, \rC);

\foreach \x in {2.5, 6, 9.5} {
    \draw[wgcore, line width=1.2pt, -Stealth] (\x-0.1, \rA) -- (\x+0.1, \rA); 
    \draw[wgcore, line width=1.2pt, -Stealth] (\x+0.1, \rB) -- (\x-0.1, \rB); 
}
\foreach \x in {6, 9.5} {
    \draw[wgcore, line width=1.2pt, -Stealth] (\x-0.1, \rC) -- (\x+0.1, \rC); 
}

\node[anchor=east, font=\bfseries, color=wgcore] at (-2.7, \rA) {Optical In};
\node[anchor=west, font=\bfseries, color=wgcore] at (\wgR+2.7, \rC) {Optical Out};

\node[font=\Huge\bfseries, color=fadedtxt] at (6, -4.0) {\vdots};
\node[font=\normalsize\bfseries, color=fadedtxt, align=center] at (6, -5.0) {Continuous scaling to $N$ rows $\times$ $M$ chiplets};

\newcommand{\drawchip}[5]{
    \def\yoffset{#3 * 0.45} 
    \def\cyoffset{#3 * 1.6} 

    \draw[#5, line width=1.5pt] (#1, #2 + \yoffset) circle (0.32cm);
    \draw[wgclad, line width=0.5pt, fill=#5!10] (#1, #2 + \yoffset) circle (0.2cm);

    \draw[wgclad, line width=4pt] (#1, #2 + \yoffset + #3*0.3) -- (#1, #2 + \cyoffset - #3*0.4);
    \draw[#5, line width=1.5pt, -Stealth] (#1, #2 + \yoffset + #3*0.3) -- (#1, #2 + \cyoffset - #3*0.4);

    \node[rectangle, rounded corners=6pt, draw=chipborder, fill=chipfill, text=chiptxt,
          minimum width=2.4cm, minimum height=1.6cm, align=center, line width=1.2pt,
          font=\sffamily\small\bfseries]
          at (#1, #2 + \cyoffset) {Chiplet\\#4};
}

\drawchip{2}{\rA}{1}{1}{lam1}
\drawchip{6}{\rA}{1}{2}{lam2}
\drawchip{10}{\rA}{1}{3}{lam3}

\drawchip{10}{\rB}{-1}{4}{lam4}
\drawchip{6}{\rB}{-1}{5}{lam5}
\drawchip{2}{\rB}{-1}{6}{lam6}

\drawchip{2}{\rC}{1}{$n-2$}{lam7}
\drawchip{6}{\rC}{1}{$n-1$}{lam8}
\drawchip{10}{\rC}{1}{$n$}{lam9}

\end{tikzpicture}

%% file: Figures/01_Fig__GoogLeNet_V00.tex
\begin{tikzpicture}[
    >=Stealth,
    font=\scriptsize\sffamily,
    box/.style={
        draw, rectangle, rounded corners=2pt, align=center,
        text=black, inner sep=2pt, minimum height=0.7cm, minimum width=1.4cm
    },
    conv/.style={box, fill=convcolor},
    pool/.style={box, fill=poolcolor},
    concat/.style={box, fill=concatcolor, minimum width=2.8cm},
    fc/.style={box, fill=fccolor},
    act/.style={box, fill=actcolor, text=black},
    io/.style={draw, align=center, fill=white, inner sep=4pt}, 
    arrow/.style={->, thick, color=black},
    incblock/.style={box, fill=boxblue, text=white, minimum height=0.8cm, minimum width=2.8cm, font=\bfseries}
]

\def\vsep{0.6cm} 

\node[io, ellipse] (input) {input};

\node[rectangle, inner sep=0pt, thick, below left=-7cm and 1.8cm of input] (real_pic) {%
    \rotatebox{90}{\includegraphics[width=7cm, height=7cm]{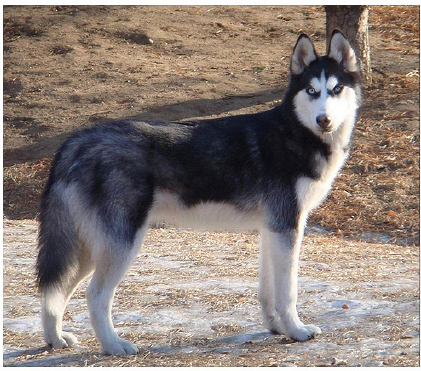}}%
};
\draw[arrow] (real_pic.east) -- (input.north west);

\node[conv, above=\vsep of input] (conv1) {Conv\\$7\times7+2(S)$};
\node[pool, above=\vsep of conv1] (pool1) {MaxPool\\$3\times3+2(S)$};
\node[concat, above=\vsep of pool1] (lrn1) {LocalRespNorm};
\node[conv, above=\vsep of lrn1] (conv2a) {Conv\\$1\times1+1(V)$};
\node[conv, above=\vsep of conv2a] (conv2b) {Conv\\$3\times3+1(S)$};
\node[concat, above=\vsep of conv2b] (lrn2) {LocalRespNorm};
\node[pool, above=\vsep of lrn2] (pool2) {MaxPool\\$3\times3+2(S)$};

\draw[arrow] (input) -- (conv1);
\draw[arrow] (conv1) -- (pool1);
\draw[arrow] (pool1) -- (lrn1);
\draw[arrow] (lrn1) -- (conv2a);
\draw[arrow] (conv2a) -- (conv2b);
\draw[arrow] (conv2b) -- (lrn2);
\draw[arrow] (lrn2) -- (pool2);

\node[incblock, above=\vsep of pool2] (inc3a) {Inception 3a};
\node[incblock, above=\vsep of inc3a] (inc3b) {Inception 3b};
\node[pool, above=\vsep of inc3b] (pool3) {MaxPool\\$3\times3+2(S)$};

\node[incblock, above=\vsep of pool3] (inc4a) {Inception 4a};
\node[incblock, above=\vsep of inc4a] (inc4b) {Inception 4b};
\node[incblock, above=\vsep of inc4b] (inc4c) {Inception 4c};
\node[incblock, above=\vsep of inc4c] (inc4d) {Inception 4d};
\node[incblock, above=\vsep of inc4d] (inc4e) {Inception 4e};
\node[pool, above=\vsep of inc4e] (pool4) {MaxPool\\$3\times3+2(S)$};

\node[incblock, above=\vsep of pool4] (inc5a) {Inception 5a};
\node[incblock, above=\vsep of inc5a] (inc5b) {Inception 5b};

\draw[arrow] (pool2) -- (inc3a);
\draw[arrow] (inc3a) -- (inc3b);
\draw[arrow] (inc3b) -- (pool3);
\draw[arrow] (pool3) -- (inc4a);
\draw[arrow] (inc4a) -- (inc4b);
\draw[arrow] (inc4b) -- (inc4c);
\draw[arrow] (inc4c) -- (inc4d);
\draw[arrow] (inc4d) -- (inc4e);
\draw[arrow] (inc4e) -- (pool4);
\draw[arrow] (pool4) -- (inc5a);
\draw[arrow] (inc5a) -- (inc5b);

\node[io, left=3.5cm of inc4e, rounded corners] (exp_in) {Previous Layer};

\def\hd{1.8} 
\def\vd{1.1} 

\node[conv] (e_1x1)    at ($(exp_in.north) + (-1.5*\hd, \vd)$) {Conv\\$1\times1$};
\node[conv] (e_3x3_r)  at ($(exp_in.north) + (-0.5*\hd, \vd)$) {Conv\\$1\times1$};
\node[conv] (e_5x5_r)  at ($(exp_in.north) + (0.5*\hd, \vd)$) {Conv\\$1\times1$};
\node[pool] (e_pool)   at ($(exp_in.north) + (1.5*\hd, \vd)$) {MaxPool\\$3\times3$};

\node[conv] (e_3x3)    at ($(e_3x3_r.north) + (0, \vd)$) {Conv\\$3\times3$};
\node[conv] (e_5x5)    at ($(e_5x5_r.north) + (0, \vd)$) {Conv\\$5\times5$};
\node[conv] (e_pool_p) at ($(e_pool.north) + (0, \vd)$) {Conv\\$1\times1$};

\node[concat, text=white] (e_concat) at ($(exp_in.north) + (0, 3.8*\vd)$) {DepthConcat};

\path ($(exp_in.north) + (0, 0.4*\vd)$) coordinate (e_div_y);
\draw[arrow] (exp_in.north) -- (exp_in.north |- e_div_y) -| (e_1x1.south);
\draw[arrow] (exp_in.north) -- (exp_in.north |- e_div_y) -| (e_3x3_r.south);
\draw[arrow] (exp_in.north) -- (exp_in.north |- e_div_y) -| (e_5x5_r.south);
\draw[arrow] (exp_in.north) -- (exp_in.north |- e_div_y) -| (e_pool.south);

\draw[arrow] (e_3x3_r.north) -- (e_3x3.south);
\draw[arrow] (e_5x5_r.north) -- (e_5x5.south);
\draw[arrow] (e_pool.north) -- (e_pool_p.south);

\path ($(exp_in.north) + (0, 3.0*\vd)$) coordinate (e_merge_y);
\draw[arrow] (e_1x1.north) -- (e_1x1.north |- e_merge_y) -| ($(e_concat.south) + (-1.0, 0)$);
\draw[arrow] (e_3x3.north) -- (e_3x3.north |- e_merge_y) -| ($(e_concat.south) + (-0.35, 0)$);
\draw[arrow] (e_5x5.north) -- (e_5x5.north |- e_merge_y) -| ($(e_concat.south) + (0.35, 0)$);
\draw[arrow] (e_pool_p.north) -- (e_pool_p.north |- e_merge_y) -| ($(e_concat.south) + (1.0, 0)$);

\begin{scope}[on background layer]
    \node[draw=boxblue, thick, dashed, rounded corners, inner xsep=10pt, inner ysep=8pt, fit=(e_1x1) (e_pool) (e_concat) (exp_in)] (blueprint_box) {};
\end{scope}
\node[font=\large\bfseries, fill=white, inner sep=2pt, below=-6.3cm of blueprint_box] {Inception Module Blueprint};

\draw[<-, dashed, thick, color=boxblue] (blueprint_box.east) -- (inc5b.west) node[midway, above, text=black, font=\bfseries] {};

\node[pool, above=\vsep of inc5b] (pool5) {AveragePool\\$7\times7+1(V)$};
\node[fc, above=\vsep of pool5] (fc_top) {FC};
\node[act, above=\vsep of fc_top] (act_top) {SoftmaxActivation};
\node[io, regular polygon, regular polygon sides=6, inner sep=1pt, above=\vsep of act_top] (out_top) {softmax2};

\draw[arrow] (inc5b) -- (pool5);
\draw[arrow] (pool5) -- (fc_top);
\draw[arrow] (fc_top) -- (act_top);
\draw[arrow] (act_top) -- (out_top);

\begin{scope}[on background layer]
    \node[draw=boxpurple, thick, dashed, rounded corners, inner sep=8pt, fit=(pool5) (out_top)] (main_box) {};
\end{scope}
\node[right=-3.2cm of main_box, font=\large\bfseries, draw=boxpurple, thick, fill=white, inner sep=6pt, align=center, rotate = 90] {Main\\Classifier};

\node[pool] (aux1_pool) at ($(inc4a.east) + (-5, -6)$) {AveragePool\\$5\times5+3(V)$};
\node[conv, above=0.4cm of aux1_pool] (aux1_conv) {Conv\\$1\times1+1(S)$};
\node[fc, above=0.4cm of aux1_conv] (aux1_fc1) {FC};
\node[fc, above=0.4cm of aux1_fc1] (aux1_fc2) {FC};
\node[act, above=0.4cm of aux1_fc2] (aux1_act) {SoftmaxActivation};
\node[io, regular polygon, regular polygon sides=6, inner sep=1pt, above=0.4cm of aux1_act] (aux1_out) {softmax0};

\draw[arrow] (inc4a.west) 
    -- ++(-0.5, 0)                        
    |- ($(aux1_pool.south) + (0, -1.0)$) 
    -- (aux1_pool.south);                

\draw[arrow] (aux1_pool) -- (aux1_conv);
\draw[arrow] (aux1_conv) -- (aux1_fc1);
\draw[arrow] (aux1_fc1) -- (aux1_fc2);
\draw[arrow] (aux1_fc2) -- (aux1_act);
\draw[arrow] (aux1_act) -- (aux1_out);

\begin{scope}[on background layer]
    \node[draw=boxorange, thick, dashed, rounded corners, inner sep=8pt, fit=(aux1_pool) (aux1_out)] (aux1_box) {};
\end{scope}

\node[pool] (aux2_pool) at ($(inc4d.east) + (-8, -8)$) {AveragePool\\$5\times5+3(V)$};
\node[conv, above=0.4cm of aux2_pool] (aux2_conv) {Conv\\$1\times1+1(S)$};
\node[fc, above=0.4cm of aux2_conv] (aux2_fc1) {FC};
\node[fc, above=0.4cm of aux2_fc1] (aux2_fc2) {FC};
\node[act, above=0.4cm of aux2_fc2] (aux2_act) {SoftmaxActivation};
\node[io, regular polygon, regular polygon sides=6, inner sep=1pt, above=0.4cm of aux2_act] (aux2_out) {softmax1};

\draw[arrow] (inc4d.west) 
    -- ++(-7.5, 0)                        
    |- ($(aux2_pool.south) + (0, -1.0)$) 
    -- (aux2_pool.south);                
    
\draw[arrow] (aux2_pool) -- (aux2_conv);
\draw[arrow] (aux2_conv) -- (aux2_fc1);
\draw[arrow] (aux2_fc1) -- (aux2_fc2);
\draw[arrow] (aux2_fc2) -- (aux2_act);
\draw[arrow] (aux2_act) -- (aux2_out);

\begin{scope}[on background layer]
    \node[draw=boxorange, thick, dashed, rounded corners, inner sep=8pt, fit=(aux2_pool) (aux2_out)] (aux2_box) {};
\end{scope}

\node[font=\Large\bfseries, draw=boxorange, thick, fill=white, inner sep=6pt, align=center] 
    at ($(aux1_box.north)!0.5!(aux2_box.north) + (-2, -9)$) (aux_label) {Auxiliary\\Classifiers};

\draw[->, color=boxorange, thick] ([xshift=0pt]aux_label.east) -- ([yshift=-40pt]aux1_box.west);
\draw[->, color=boxorange, thick] ([xshift=15pt]aux_label.north) -- ([xshift=10pt]aux2_box.south);

\begin{scope}[shift={($(blueprint_box.south) + (0, 9.5)$)}]
    \node[draw=gray, thick, rounded corners, fill=white, minimum width=5.0cm, minimum height=4.5cm] (leg_bg) at (0, -0.5) {};
    \node[font=\large\bfseries] at (0, 1.2) {Color Legend};
    
    \node[box, fill=convcolor, minimum width=0.8cm, minimum height=0.4cm] (lc) at (-1.8, 0.5) {};
    \node[right=0.2cm of lc, anchor=west, font=\normalsize] {Convolution};
    
    \node[box, fill=poolcolor, minimum width=0.8cm, minimum height=0.4cm] (lp) at (-1.8, -0.2) {};
    \node[right=0.2cm of lp, anchor=west, font=\normalsize] {Pooling};
    
    \node[box, fill=concatcolor, minimum width=0.8cm, minimum height=0.4cm] (lco) at (-1.8, -0.9) {};
    \node[right=0.2cm of lco, anchor=west, font=\normalsize] {Concat/Normalize};
    
    \node[box, fill=fccolor, minimum width=0.8cm, minimum height=0.4cm] (lfc) at (-1.8, -1.6) {};
    \node[right=0.2cm of lfc, anchor=west, font=\normalsize] {Fully Connected};
    
    \node[box, fill=actcolor, text=black, minimum width=0.8cm, minimum height=0.4cm] (la) at (-1.8, -2.3) {};
    \node[right=0.2cm of la, anchor=west, font=\normalsize] {Softmax/Activation};
\end{scope}

\end{tikzpicture}

%% file: Figures/02_ResnetFig.tex
\begin{tikzpicture}[
    >=Stealth,
    node distance=0.4cm,
    block/.style={
        draw, rectangle, minimum height=3.2cm, minimum width=0.7cm,
        fill=#1, rounded corners=2pt, font=\scriptsize,
        inner sep=3pt, align=center, text=white
    },
    poolblock/.style={
        draw, rectangle, minimum height=1.4cm, minimum width=0.6cm,
        fill=poolgray, rounded corners=2pt, font=\scriptsize,
        inner sep=2pt, align=center
    },
    arrow/.style={->, thick, color=black},
    curvedarrow/.style={->, thick, color=arrowblue, line width=1.2pt},
    sizelabel/.style={font=\tiny, text=black, rotate=90},
]

\node[rectangle, inner sep=0pt] (input) {%
    \includegraphics[height=1.8cm, width=1.8cm]{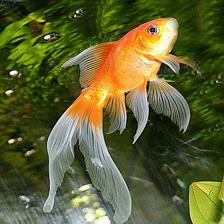}%
};

\node[block=convgreen, right=0.6cm of input, minimum width=0.7cm, minimum height=2.4cm] (conv1) {%
    \rotatebox{90}{7$\times$7 Conv, 64/2}};

\node[poolblock, right=0.5cm of conv1] (pool1) {\rotatebox{90}{2$\times$2 Pool}};

\node[block=blockblue, right=0.7cm of pool1] (b1c) {%
    \rotatebox{90}{\shortstack{1$\times$1 Conv,64\\3$\times$3 Conv,64\\1$\times$1 Conv,256}}};
\node[block=blockblue] at ($(b1c)+(-0.10,-0.10)$) (b1b) {%
    \rotatebox{90}{\shortstack{1$\times$1 Conv,64\\3$\times$3 Conv,64\\1$\times$1 Conv,256}}};
\node[block=blockblue] at ($(b1c)+(-0.20,-0.20)$) (b1a) {%
    \rotatebox{90}{\shortstack{1$\times$1 Conv,64\\3$\times$3 Conv,64\\1$\times$1 Conv,256}}};
\node[font=\tiny] at ($(b1a.north west)!0.5!(b1c.north east)+(0.1,0.8)$) {\textbf{$\times$3}};

\draw[curvedarrow]
    ($(b1a.north west)+(0.3, 0.25)$)
    .. controls +(0, 0.7) and +(0, 0.7) ..
    ($(b1c.north east)+(-0.05, 0.0)$);

\node[block=blockblue, right=1.0cm of b1c] (b2c) {%
    \rotatebox{90}{\shortstack{1$\times$1 Conv,128/2\\3$\times$3 Conv,128\\1$\times$1 Conv,512}}};
\node[block=blockblue] at ($(b2c)+(-0.10,-0.10)$) (b2b) {%
    \rotatebox{90}{\shortstack{1$\times$1 Conv,128/2\\3$\times$3 Conv,128\\1$\times$1 Conv,512}}};
\node[block=blockblue] at ($(b2c)+(-0.20,-0.20)$) (b2a) {%
    \rotatebox{90}{\shortstack{1$\times$1 Conv,128/2\\3$\times$3 Conv,128\\1$\times$1 Conv,512}}};
\node[font=\tiny] at ($(b2a.north west)!0.5!(b2c.north east)+(0.1,0.8)$) {\textbf{$\times$3}};

\draw[curvedarrow]
    ($(b2a.north west)+(0.3, 0.25)$)
    .. controls +(0, 0.7) and +(0, 0.7) ..
    ($(b2c.north east)+(-0.05, 0.0)$);

\node[block=blockblue, right=1.0cm of b2c] (b3c) {%
    \rotatebox{90}{\shortstack{1$\times$1 Conv,256/2\\3$\times$3 Conv,256\\1$\times$1 Conv,1024}}};
\node[block=blockblue] at ($(b3c)+(-0.10,-0.10)$) (b3b) {%
    \rotatebox{90}{\shortstack{1$\times$1 Conv,256/2\\3$\times$3 Conv,256\\1$\times$1 Conv,1024}}};
\node[block=blockblue] at ($(b3c)+(-0.20,-0.20)$) (b3a) {%
    \rotatebox{90}{\shortstack{1$\times$1 Conv,256/2\\3$\times$3 Conv,256\\1$\times$1 Conv,1024}}};
\node[font=\tiny] at ($(b3a.north west)!0.5!(b3c.north east)+(0.1,0.8)$) {\textbf{$\times$3}};

\draw[curvedarrow]
    ($(b3a.north west)+(0.3, 0.25)$)
    .. controls +(0, 0.7) and +(0, 0.7) ..
    ($(b3c.north east)+(-0.05, 0.0)$);

\node[block=blockblue, right=1.0cm of b3c] (b4c) {%
    \rotatebox{90}{\shortstack{1$\times$1 Conv,512/2\\3$\times$3 Conv,512\\1$\times$1 Conv,2048}}};
\node[block=blockblue] at ($(b4c)+(-0.10,-0.10)$) (b4b) {%
    \rotatebox{90}{\shortstack{1$\times$1 Conv,512/2\\3$\times$3 Conv,512\\1$\times$1 Conv,2048}}};
\node[block=blockblue] at ($(b4c)+(-0.20,-0.20)$) (b4a) {%
    \rotatebox{90}{\shortstack{1$\times$1 Conv,512/2\\3$\times$3 Conv,512\\1$\times$1 Conv,2048}}};
\node[font=\tiny] at ($(b4a.north west)!0.5!(b4c.north east)+(0.1,0.8)$) {\textbf{$\times$3}};

\draw[curvedarrow]
    ($(b4a.north west)+(0.3, 0.25)$)
    .. controls +(0, 0.7) and +(0, 0.7) ..
    ($(b4c.north east)+(-0.05, 0.0)$);

\node[poolblock, right=1.0cm of b4c, minimum height=1.6cm] (pool2) {\rotatebox{90}{Avg Pool}};

\node[block=fccolor, right=0.6cm of pool2, minimum width=0.7cm, minimum height=2.4cm] (fc)
    {\rotatebox{90}{FC 1000}};

\node[right=0.6cm of fc] (softmax_label) {\scriptsize Softmax};

\foreach \yshift/\idx in {0.7/1, 0.35/2, -0.35/4, -0.7/5} {
    \node[draw, circle, minimum size=0.25cm, fill=white]
        at ($(softmax_label.east)+(1.0,\yshift)$) (out\idx) {};
}
\node at ($(softmax_label.east)+(1.0,0)$) {$\vdots$};
\node[font=\scriptsize, rotate=90, anchor=south]
    at ($(softmax_label.east)+(1.6,0)$) {Outputs (1$\times$ nb\_personID)};

\node[font=\tiny, rotate=90, anchor=center]
    at ($(input.east)!0.5!(conv1.west)+(0,-1.0)$)  {Size 224};
\node[font=\tiny, rotate=90, anchor=center]
    at ($(conv1.east)!0.5!(pool1.west)+(0,-1.0)$)  {Size 112};
\node[font=\tiny, rotate=90, anchor=center]
    at ($(pool1.east)!0.5!(b1a.west)+(0,-1.0)$)    {Size 56};
\node[font=\tiny, rotate=90, anchor=center]
    at ($(b1c.east)!0.8!(b2a.west)+(0,-1.0)$)      {Size 28};
\node[font=\tiny, rotate=90, anchor=center]
    at ($(b2c.east)!0.8!(b3a.west)+(0,-1.0)$)      {Size 14};
\node[font=\tiny, rotate=90, anchor=center]
    at ($(b3c.east)!0.8!(b4a.west)+(0,-1.0)$)      {Size 7};

\draw[arrow] (input)   -- (conv1);
\draw[arrow] (conv1)   -- (pool1);
\draw[arrow] (pool1)   -- ([yshift=6pt]b1a.west);
\draw[arrow] ([yshift=0pt]b1c.east) -- ([yshift=6pt]b2a.west);
\draw[arrow] ([yshift=0pt]b2c.east) -- ([yshift=6pt]b3a.west);
\draw[arrow] ([yshift=0pt]b3c.east) -- ([yshift=6pt]b4a.west);
\draw[arrow] (b4c)     -- (pool2);
\draw[arrow] (pool2)   -- (fc);
\draw[arrow] (fc)      -- (softmax_label);
\draw[arrow] (softmax_label) -- (out1);
\draw[arrow] (softmax_label) -- (out2);
\draw[arrow] (softmax_label) -- (out4);
\draw[arrow] (softmax_label) -- (out5);

\end{tikzpicture}%

%% file: ref.bib
@article{jouppi2017datacenter,
  title={In-datacenter performance analysis of a tensor processing unit},
  author={Jouppi, Norman P and Young, Cliff and Patil, Nishant and Patterson, David and Agrawal, Gaurav and Bajwa, Raminder and Bates, Sarah and Bhatia, Suresh and Boden, Nan and Borchers, Al and others},
  journal={ACM SIGARCH Computer Architecture News},
  volume={45},
  number={2},
  pages={1--12},
  year={2017},
  publisher={ACM}
}

@inproceedings{dalir2022high_density,
  title={High-density Integrated Photonic Tensor Processing Unit with a Matrix Multiply Compile},
  author={Dalir, Hamed and Nouri, Behrouz Movahhed and Ma, Xiaoxuan and Nicola, Peserico and Shastri, Bhavin J and Sorger, Volker J},
  booktitle={2022 IEEE Photonics Conference (IPC)},
  pages={1--2},
  year={2022},
  organization={IEEE}
}

@article{samajdar2020systematic,
  title={A systematic methodology for characterizing scalability of DNN accelerators using SCALE-Sim},
  author={Samajdar, Ananda and Joseph, Jan Moritz and Zhu, Yuhao and Whatmough, Paul and Mattina, Matthew and Krishna, Tushar},
  journal={IEEE Computer Architecture Letters},
  volume={19},
  number={2},
  pages={114--117},
  year={2020},
  publisher={IEEE}
}

@inproceedings{chen2016eyeriss,
  title={Eyeriss: An energy-efficient reconfigurable accelerator for deep convolutional neural networks},
  author={Chen, Yu-Hsin and Emer, Joel and Sze, Vivienne},
  booktitle={IEEE International Solid-State Circuits Conference (ISSCC)},
  pages={262--263},
  year={2016},
  organization={IEEE}
}

@article{howard2017mobilenets,
  title={Mobilenets: Efficient convolutional neural networks for mobile vision applications},
  author={Howard, Andrew G and Zhu, Menglong and Chen, Bo and Kalenichenko, Dmitry and Wang, Weijun and Weyand, Tobias and Andreetto, Marco and Adam, Hartwig},
  journal={arXiv preprint arXiv:1704.04861},
  year={2017}
}

@inproceedings{szegedy2015going,
  title={Going deeper with convolutions},
  author={Szegedy, Christian and Liu, Wei and Jia, Yangqing and Sermanet, Pierre and Reed, Scott and Anguelov, Dragomir and Erhan, Dumitru and Vanhoucke, Vincent and Rabinovich, Andrew},
  booktitle={Proceedings of the IEEE conference on computer vision and pattern recognition},
  pages={1--9},
  year={2015}
}

@article{zhang2020photonic,
  title={Photonic neural networks: a survey},
  author={Zhang, Weiwen and Yao, Yichen and others},
  journal={IEEE Journal of Selected Topics in Quantum Electronics},
  volume={26},
  number={5},
  pages={1--12},
  year={2020},
  publisher={IEEE}
}

@article{gholami2024aia,
  title={Ai and memory wall},
  author={Gholami, Amir and Yao, Zhewei and Kim, Sehoon and Hooper, Coleman and Mahoney, Michael W and Keutzer, Kurt},
  journal={IEEE Micro},
  volume={44},
  number={3},
  pages={33--39},
  year={2024},
  publisher={IEEE}
}

@article{jouppi2018motivation,
  title={Motivation for and evaluation of the first tensor processing unit},
  author={Jouppi, Norman and Young, Cliff and Patil, Nishant and Patterson, David},
  journal={ieee Micro},
  volume={38},
  number={3},
  pages={10--19},
  year={2018},
  publisher={IEEE}
}

@article{zhou2022photonic,
  title={Photonic matrix multiplication lights up photonic accelerator and beyond},
  author={Zhou, Hailong and Dong, Jianji and Cheng, Junwei and Dong, Wenchan and Huang, Chaoran and Shen, Yichen and Zhang, Qiming and Gu, Min and Qian, Chao and Chen, Hongsheng and others},
  journal={Light: Science \& Applications},
  volume={11},
  number={1},
  pages={30},
  year={2022},
  publisher={Nature Publishing Group UK London}
}

@article{hua2025integrated,
  title={An integrated large-scale photonic accelerator with ultralow latency},
  author={Hua, Shiyue and Divita, Erwan and Yu, Shanshan and Peng, Bo and Roques-Carmes, Charles and Su, Zhan and Chen, Zhang and Bai, Yanfei and Zou, Jinghui and Zhu, Yunpeng and others},
  journal={Nature},
  volume={640},
  number={8058},
  pages={361--367},
  year={2025},
  publisher={Nature Publishing Group UK London}
}

@article{ahmed2025universal,
  title={Universal photonic artificial intelligence acceleration},
  author={Ahmed, Sufi R and Baghdadi, Reza and Bernadskiy, Mikhail and Bowman, Nate and Braid, Ryan and Carr, Jim and Chen, Chen and Ciccarella, Pietro and Cole, Matthew and Cooke, John and others},
  journal={Nature},
  volume={640},
  number={8058},
  pages={368--374},
  year={2025},
  publisher={Nature Publishing Group UK London}
}

@article{al2022scaling,
  title={Scaling up silicon photonic-based accelerators: Challenges and opportunities},
  author={Al-Qadasi, MA and Chrostowski, L and Shastri, BJ and Shekhar, S},
  journal={APL Photonics},
  volume={7},
  number={2},
  year={2022},
  publisher={AIP Publishing}
}

@inproceedings{jahannia2024low,
  title={Low-latency full precision optical convolutional neural network accelerator},
  author={Jahannia, Belal and Ye, Jiachi and Altaleb, Salem and Peserico, Nicola and Asadizanjani, Navid and Heidari, Elham and Sorger, Volker J and Dalir, Hamed},
  booktitle={AI and Optical Data Sciences V},
  volume={12903},
  pages={27--41},
  year={2024},
  organization={SPIE}
}

@inproceedings{yin2024scatter,
  title={Scatter: algorithm-circuit co-sparse photonic accelerator with thermal-tolerant, power-efficient in-situ light redistribution},
  author={Yin, Ziang and Gangi, Nicholas and Zhang, Meng and Zhang, Jeff and Huang, Rena and Gu, Jiaqi},
  booktitle={Proceedings of the 43rd IEEE/ACM International Conference on Computer-Aided Design},
  pages={1--9},
  year={2024}
}

@inproceedings{he2016deep,
  title={Deep residual learning for image recognition},
  author={He, Kaiming and Zhang, Xiangyu and Ren, Shaoqing and Sun, Jian},
  booktitle={Proceedings of the IEEE conference on computer vision and pattern recognition},
  pages={770--778},
  year={2016}
}

@article{deng2012mnist,
  title={The mnist database of handwritten digit images for machine learning research [best of the web]},
  author={Deng, Li},
  journal={IEEE signal processing magazine},
  volume={29},
  number={6},
  pages={141--142},
  year={2012},
  publisher={IEEE}
}

@article{feldmann2021parallel,
  title={Parallel convolutional processing using an integrated photonic tensor core},
  author={Feldmann, Johannes and Youngblood, Nathan and Karpov, Maxim and Gehring, Helge and Li, Xuan and Stappers, Maik and Le Gallo, Manuel and Fu, Xin and Lukashchuk, Anton and Raja, Arslan S and others},
  journal={Nature},
  volume={589},
  number={7840},
  pages={52--58},
  year={2021},
  publisher={Nature Publishing Group UK London}
}

@article{ma2022high,
  title={High-density integrated photonic tensor processing unit with a matrix multiply compiler},
  author={Ma, Xiaoxuan and Peserico, Nicola and Khaled, Ahmed and Guo, Zhimo and Nouri, Behrouz and Dalir, Hamed and Shastri, Bhavin and Sorger, Volker},
  Journal = {Research Square},
  year={2022}
}

@article{ning2025hardware,
  title={Hardware-efficient photonic tensor core: accelerating deep neural networks with structured compression},
  author={Ning, Shupeng and Zhu, Hanqing and Feng, Chenghao and Gu, Jiaqi and Pan, David Z and Chen, Ray T},
  journal={Optica},
  volume={12},
  number={7},
  pages={1079--1089},
  year={2025},
  publisher={Optica Publishing Group}
}

@article{dong2023higher,
  title={Higher-dimensional processing using a photonic tensor core with continuous-time data},
  author={Dong, Bowei and Aggarwal, Samarth and Zhou, Wen and Ali, Utku Emre and Farmakidis, Nikolaos and Lee, June Sang and He, Yuhan and Li, Xuan and Kwong, Dim-Lee and Wright, CD and others},
  journal={Nature Photonics},
  volume={17},
  number={12},
  pages={1080--1088},
  year={2023},
  publisher={Nature Publishing Group UK London}
}

@article{silver2017mastering,
  title={Mastering the game of go without human knowledge},
  author={Silver, David and Schrittwieser, Julian and Simonyan, Karen and Antonoglou, Ioannis and Huang, Aja and Guez, Arthur and Hubert, Thomas and Baker, Lucas and Lai, Matthew and Bolton, Adrian and others},
  journal={nature},
  volume={550},
  number={7676},
  pages={354--359},
  year={2017},
  publisher={Nature Publishing Group UK London}
}

@article{farmakidis2019plasmonic,
  title={Plasmonic nanogap enhanced phase-change devices with dual electrical-optical functionality},
  author={Farmakidis, Nikolaos and Youngblood, Nathan and Li, Xuan and Tan, James and Swett, Jacob L and Cheng, Zengguang and Wright, C David and Pernice, Wolfram HP and Bhaskaran, Harish},
  journal={Science advances},
  volume={5},
  number={11},
  pages={eaaw2687},
  year={2019},
  publisher={American Association for the Advancement of Science}
}

@article{sawant2025high,
  title={High-Endurance and Energy-Efficient All-Optical Programming of Scalable Silicon Waveguides with Integrated Phase-Change Material Patches},
  author={Sawant, Rajath and Albanese, Anthony and Rogemont, Arnaud and Gonzalez-Cortes, Gregorio and Br{\^u}l{\'e}, Yoann and Karam, Lara and Jager, Jean-baptiste and Malhouitre, Stephane and Charbonnier, Benoit and Coillet, Aur{\'e}lien and others},
  journal={Advanced Optical Materials},
  volume={13},
  number={25},
  pages={e00775},
  year={2025},
  publisher={Wiley Online Library}
}

@inproceedings{shafiee2023compact,
  title={Compact and low-loss PCM-based silicon photonic MZIs for photonic neural networks},
  author={Shafiee, Amin and Banerjee, Sanmitra and Charbonnier, Benoit and Pasricha, Sudeep and Nikdast, Mahdi},
  booktitle={2023 IEEE Photonics Conference (IPC)},
  pages={1--2},
  year={2023},
  organization={IEEE}
}

@article{stepanovsky2019comparative,
  title={A comparative review of MEMS-based optical cross-connects for all-optical networks from the past to the present day},
  author={Stepanovsky, Michal},
  journal={IEEE Communications Surveys \& Tutorials},
  volume={21},
  number={3},
  pages={2928--2946},
  year={2019},
  publisher={IEEE}
}

@article{aadhi2025scalable,
  title={Scalable photonic reservoir computing for parallel machine learning tasks},
  author={Aadhi, A and Di Lauro, L and Fischer, B and Dmitriev, P and Alamgir, I and Mazoukh, C and Perron, N and Viktorov, EA and Kovalev, AV and Eshaghi, A and others},
  journal={Nature Communications},
  year={2025},
  publisher={Nature Publishing Group UK London}
}

@article{shekhar2024roadmapping,
  title={Roadmapping the next generation of silicon photonics},
  author={Shekhar, Sudip and Bogaerts, Wim and Chrostowski, Lukas and Bowers, John E and Hochberg, Michael and Soref, Richard and Shastri, Bhavin J},
  journal={Nature Communications},
  volume={15},
  number={1},
  pages={751},
  year={2024},
  publisher={Nature Publishing Group UK London}
}

@inproceedings{tian2025photonic,
  title={Photonic integrated circuits: research advances and challenges in interconnection and packaging technologies},
  author={Tian, Wenchao and Wang, Yifan and Dang, Haojie and Hou, Huahua and Xi, Yuanyuan},
  booktitle={Photonics},
  volume={12},
  number={8},
  pages={821},
  year={2025},
  organization={MDPI}
}

@article{wu2020state,
  title={State-of-the-art and perspectives on silicon waveguide crossings: A review},
  author={Wu, Sailong and Mu, Xin and Cheng, Lirong and Mao, Simei and Fu, HY},
  journal={Micromachines},
  volume={11},
  number={3},
  pages={326},
  year={2020},
  publisher={MDPI}
}

@inproceedings{tu202228nm,
  title={A 28nm 29.2 TFLOPS/W BF16 and 36.5 TOPS/W INT8 reconfigurable digital CIM processor with unified FP/INT pipeline and bitwise in-memory booth multiplication for cloud deep learning acceleration},
  author={Tu, Fengbin and Wang, Yiqi and Wu, Zihan and Liang, Ling and Ding, Yufei and Kim, Bongjin and Liu, Leibo and Wei, Shaojun and Xie, Yuan and Yin, Shouyi},
  booktitle={2022 IEEE International Solid-State Circuits Conference (ISSCC)},
  volume={65},
  pages={1--3},
  year={2022},
  organization={IEEE}
}

@article{le202364,
  title={A 64-core mixed-signal in-memory compute chip based on phase-change memory for deep neural network inference},
  author={Le Gallo, Manuel and Khaddam-Aljameh, Riduan and Stanisavljevic, Milos and Vasilopoulos, Athanasios and Kersting, Benedikt and Dazzi, Martino and Karunaratne, Geethan and Br{\"a}ndli, Matthias and Singh, Abhairaj and Mueller, Silvia M and others},
  journal={Nature Electronics},
  volume={6},
  number={9},
  pages={680--693},
  year={2023},
  publisher={Nature Publishing Group UK London}
}

@article{yuzuguler2023scale,
  title={Scale-out systolic arrays},
  author={Y{\"u}z{\"u}g{\"u}ler, Ahmet Caner and S{\"o}nmez, Canberk and Drumond, Mario and Oh, Yunho and Falsafi, Babak and Frossard, Pascal},
  journal={ACM Transactions on Architecture and Code Optimization},
  volume={20},
  number={2},
  pages={1--25},
  year={2023},
  publisher={ACM New York, NY}
}

@article{tait2014broadcast,
  title={Broadcast and weight: an integrated network for scalable photonic spike processing},
  author={Tait, Alexander N and Nahmias, Mitchell A and Shastri, Bhavin J and Prucnal, Paul R},
  journal={Journal of Lightwave Technology},
  volume={32},
  number={21},
  pages={3427--3439},
  year={2014},
  publisher={OSA}
}

@article{kim2024photonic,
  title={Photonic Systolic Array for All-Optical Matrix--Matrix Multiplication},
  author={Kim, Jungmin and Zhou, Qingyi and Yu, Zongfu},
  journal={Laser \& Photonics Reviews},
  pages={e01995},
  year={2024},
  publisher={Wiley Online Library}
}

@article{tang2025waveguide,
  title={Waveguide-multiplexed photonic matrix--vector multiplication processor using multiport photodetectors},
  author={Tang, Rui and Okano, Makoto and Zhang, Chao and Toprasertpong, Kasidit and Takagi, Shinichi and Takenaka, Mitsuru},
  journal={Optica},
  volume={12},
  number={6},
  pages={812--820},
  year={2025},
  publisher={Optica Publishing Group}
}

@inproceedings{raj2025scale,
  title={SCALE-Sim v3: A modular cycle-accurate systolic accelerator simulator for end-to-end system analysis},
  author={Raj, Ritik and Banerjee, Sarbartha and Chandra, Nikhil and Wan, Zishen and Tong, Jianming and Samajdhar, Ananda and Krishna, Tushar},
  booktitle={2025 IEEE International Symposium on Performance Analysis of Systems and Software (ISPASS)},
  pages={186--200},
  year={2025},
  organization={IEEE}
}

@article{song2025integrated,
  title={Integrated electro-optic digital-to-analogue link for efficient computing and arbitrary waveform generation},
  author={Song, Yunxiang and Hu, Yaowen and Zhu, Xinrui and Powell, Keith and Magalh{\~a}es, Let{\'\i}cia and Ye, Fan and Warner, Hana K and Lu, Shengyuan and Li, Xudong and Renaud, Dylan and others},
  journal={Nature Photonics},
  volume={19},
  number={10},
  pages={1107--1115},
  year={2025},
  publisher={Nature Publishing Group UK London}
}

@inproceedings{cai2025tcsel,
  title={TCSEL: a novel high-power laser architecture with enhanced efficiency and beam control for next-generation LiDAR applications},
  author={Cai, Qian and Jahannia, Belal and Wu, Tongyao and Ye, Jiachi and Amirany, Abdolah and Dalir, Hamed and Heidari, Elham},
  booktitle={Infrared Sensors, Devices, and Applications XV},
  volume={13613},
  pages={105--114},
  year={2025},
  organization={SPIE}
}

@inproceedings{amirany2025integrated,
  title={Integrated PCM-based DAC/ADC design for energy-efficient photonic computing systems},
  author={Amirany, Abdolah and Jahannia, Belal and Hayden, Benjamin and Karkheiran, Kasra and Heidari, Elham and Dalir, Hamed},
  booktitle={Photonic Computing: From Materials and Devices to Systems and Applications II},
  volume={13581},
  pages={51--61},
  year={2025},
  organization={SPIE}
}

@inproceedings{wu2025highly,
  title={Highly Energy-efficient Photonic Modulator Based on Exceptional Point},
  author={Wu, Tongyao and Cai, Qian and Ye, Jiachi and Jahannia, Belal and Liao, Ming-Han and Dalir, Hamed and Heidari, Elham},
  booktitle={CLEO: Science and Innovations},
  pages={JPS100\_169},
  year={2025},
  organization={Optica Publishing Group}
}

@article{tossoun2025large,
  title={Large-scale integrated photonic device platform for energy-efficient AI/ML accelerators},
  author={Tossoun, Bassem and Xiao, Xian and Cheung, Stanley and Yuan, Yuan and Peng, Yiwei and Srinivasan, Sudharsanan and Giamougiannis, George and Huang, Zhihong and Singaraju, Prerana and London, Yanir and others},
  journal={IEEE Journal of Selected Topics in Quantum Electronics},
  volume={31},
  number={3: AI/ML Integrated Opto-electronics},
  pages={1--26},
  year={2025},
  publisher={IEEE}
}

@article{parra2024silicon,
  title={Silicon thermo-optic phase shifters: a review of configurations and optimization strategies},
  author={Parra, Jorge and Navarro-Arenas, Juan and Sanchis, Pablo},
  journal={Advanced Photonics Nexus},
  volume={3},
  number={4},
  pages={044001--044001},
  year={2024},
  publisher={Society of Photo-Optical Instrumentation Engineers}
}
